\documentclass[fleqn,usenatbib]{mnras}

\usepackage{newtxtext,newtxmath}
\usepackage{multicol}
\usepackage{graphicx}

\usepackage[T1]{fontenc}

\DeclareRobustCommand{\VAN}[3]{#2}
\let\VANthebibliography\thebibliography
\def\thebibliography{\DeclareRobustCommand{\VAN}[3]{##3}\VANthebibliography}

\usepackage{graphicx}	% Including figure files
\usepackage{amsmath}	% Advanced maths commands

\usepackage{deluxetable}
\usepackage{float}
\usepackage{caption}
\usepackage{bm}
\title[Abundances for a ``hot Saturn'' exoplanet]{Solar-to-supersolar sodium and oxygen absolute abundances for \\ a ``hot Saturn'' orbiting a metal-rich star}
\author[N. Nikolov et al.]{Nikolay K. Nikolov,$^{1}$\thanks{E-mail: nnikolov@stsci.edu (NN)}
David K. Sing,$^{2,3}$
Jessica J. Spake,$^{4}$
Barry Smalley,$^{5}$
Jayesh M. Goyal,$^{6}$
\newauthor 
Thomas Mikal-Evans,$^{7}$
Hannah R. Wakeford,$^{8}$
Zafar Rustamkulov,$^{3}$
Drake Deming,$^{9}$
\newauthor 
Jonathan J. Fortney,$^{10}$
Aarynn Carter, $^{10}$
Neale P. Gibson,$^{11}$
Nathan J. Mayne$^{12}$
\\
$^{1}$Space Telescope Science Institute, 3700 San Martin Dr, Baltimore, MD 21218, USA\\
$^{2}$Department of Physics \& Astronomy, Johns Hopkins University, Baltimore, MD 21210, USA\\
$^{3}$Department of Earth \& Planetary Sciences, Johns Hopkins University, Baltimore, MD 21210, USA\\
$^{4}$Division of Geological and Planetary Sciences, California Institute of Technology, 1200 East California Blvd, Pasadena, CA 91125, USA\\
$^{5}$Astrophysics Group, Keele University,  Newcastle-under-Lyme, Staffordshire, ST5 5BG, UK\\
$^{6}$National Institute of Science Education and Research (NISER), HBNI,
Jatni, Khurda-752050, Odisha, India\\
$^{7}$Max-Planck Institute for Astronomy, K\"{o}nigstuhl 17, Heidelberg, 69117, Germany\\
$^{8}$School of Physics, University of Bristol, HH Wills Physics Laboratory, Tyndall Avenue, Bristol BS8 1TL, UK\\
$^{9}$Department of Astronomy, University of Maryland, College Park, MD 20742, USA\\
$^{10}$Department of Astronomy and Astrophysics, University of California, Santa Cruz, Santa Cruz, CA 95064, USA\\
$^{11}$School of Physics, Trinity College Dublin, Dublin 2, Ireland\\
$^{12}$Department of Astrophysics College of Engineering, Mathematics, and Physical Sciences, University of Exeter, Exeter, EX4 4QF, UK\\
}

\date{Accepted 2022 May 30. Received 2022 May 30; in original form 2022 March 14}

\pubyear{2021}

\begin{document}
\label{firstpage}
\pagerange{\pageref{firstpage}--\pageref{lastpage}}
\maketitle

% $\mathrm{[O/H]}=0.87^{+0.38}_{-0.43}$  and $\mathrm{[Na/H]}=1.32^{+0.36}_{-0.46}$


%  Using the {\tt{ATMO}} retrieval code, and assuming chemical equilibrium, w
% Comparisons of elemental abundances from exoplanets with cloud-free atmospheres with the abundances of their host stars may shed light on their formation and evolutionary history in the Solar System and wider galactic context.

%Comparisons of the elemental abundances inferred from observations of cloud-free atmospheres of exoplanets to those derived for their host stars, may provide important insights into the formation and evolution of planets. 

\begin{abstract}
We present new analysis of infrared transmission spectroscopy of the cloud-free hot-Saturn WASP-96b performed with the {\it{Hubble}} and {\it{Spitzer Space Telescopes}} ({\it{HST}} and {\it{Spitzer}}). The WASP-96b spectrum exhibits the absorption feature from water in excellent agreement with synthetic spectra computed assuming a cloud-free atmosphere. The {\it{HST-Spitzer}} spectrum is coupled with Very Large Telescope (VLT) optical transmission spectroscopy which reveals the full pressure-broadened profile of the sodium absorption feature and enables the derivation of absolute abundances. We confirm and correct for a spectral offset of $\Delta R_{{\rm p}}/R_{\ast}=(-4.29^{+0.31}_{-0.37})\,\times10^{-3}$ of the VLT data relative to the {\it{HST-Spitzer}} spectrum. This offset can be explained by the assumed radius for the common-mode correction of the VLT spectra, which is a well-known feature of ground-based transmission spectroscopy. We find evidence for a lack of chromospheric and photometric activity of the host star which, therefore, make a negligible contribution to the offset. We measure abundances for Na and O that are consistent with solar to supersolar, with abundances relative to solar values of {\mbox{$21^{+27}_{-14}$ and $7^{+11}_{-4}$}}, respectively.  We complement the transmission spectrum with new thermal emission constraints from {\it{Spitzer}} observations at 3.6 and $4.5\mu$m, which are best explained by the spectrum of an atmosphere with a temperature decreasing with altitude. A fit to the spectrum assuming an isothermal blackbody atmosphere constrains the dayside temperature to be $T_{\rm{p}}$=$1545$$\pm$$90$K. 

%While metal enrichement of exoplanets has been shown not to correlate with the metallicity of their parent stars, the case of WASP-96 is a prime example for a significant enrichment of both the star and planet, which likely reflects the initial formation conditions of this system.

%With a planet-to-star oxygen abundance ratio of $\sim3\times$, which is a match to the corresponding ratio for Jupiter and the Sun, WASP-96b may have formed under similar conditions to Jupiter. 

% This makes the cloud-free atmosphere of WASP-96b a prototype for further detailed abundance constraints with the {\it{James Webb Space Telescope.}}

%Our results are in an excellent agreement with the supersolar elemental abundances for the host star, which we constrain from archival high-resolution spectroscopy.  
\end{abstract}

%Assuming a solar and supersolar abundance of the planet atmosphere, we find the measured emission spectrum can be best explained by the spectrum of an atmosphere with a decreasing temperature with altitude.

% We constrain the chromospheric activity of the host star, which we find to be low with an $\log{\rm{R^{'}_{H\&K}}}=-5.3\pm0.1$, and insufficient to explain the observed offset. 

%  which we measure and correct in our retrieval analysis

\begin{keywords}
planets and satellites: atmospheres -- stars: abundances -- techniques: spectroscopic -- methods: observational -- methods: data analysis
\end{keywords}

%%%%%%%%%%%%%%%%%%%%%%%%%%%%%%%%%%%%%%%%%%%%%%%%%%

%%%%%%%%%%%%%%%%% BODY OF PAPER %%%%%%%%%%%%%%%%%%

%Transit, eclipse and phase curve spectroscopy of irradiated gas giant exoplanets continues to provide valuable constraints of their composition, temperature and circulation. By comparing theory with observations, we inform our understanding about the processes shaping the formation and evolution of exoplanets in the Solar System and broad galactic context. 

%With masses and radii similar to Jupiter in our own Solar System, but orbital periods shorter than 10 days, hot Jupiters are among the most favourable exoplanets for atmospheric characterisation. 

\section{Introduction} \label{sec:intro}
From massive gas-giants down to Earth-mass worlds, from ultra-hot planets to much cooler and potentially habitable Earth-like analogs, exoplanets have now been found orbiting many of the stars across the solar neighborhood. Following nearly three decades of exoplanet detections, the exoplanet census now shows that planet formation is ubiquitous with Super-Earths being common and hot Jupiters, planets similar to Jupiter in the Solar System but with orbital periods shorter than about 10 days, uncommon \citep{2015ApJ...809....8B, 2015ApJ...799..229W, 2010Sci...330..653H, 2012ApJS..201...15H, 2011arXiv1109.2497M, 2016ApJS..225...32B}. Due to their large radii and hot temperatures, hot Jupiters host the only class of exoplanet atmosphere that can currently be observationally characterized in detail with spectroscopic observations over an entire orbital phase \cite{2021JGRE..12606629F}. Even the first steps to characterize exoplanets via transits have started to reveal astoundingly diverse worlds very different from the planets of the Solar System with: (i) evaporating and escaping planetary atmospheres resembling comet-like tails \citep{2015Natur.522..459E}; (ii) atmospheric hazes and clouds composed of minerals and metal oxides, which tend to be ubiquitous \citep{lecavelier08a, Sing2013, huitson13, nikolov15, 2020MNRAS.494.5449C, 2020AJ....160...51A, 2021MNRAS.506.2853S, 2020MNRAS.497.5155W}; and (iii) a huge diversity and a continuum from clear to cloudy atmospheres \citep{sing16, 2017MNRAS.467.4591G, 2018AJ....155...29W, 2018AJ....156..283E, 2021ApJ...906L..10A, 2021AJ....162..108F, 2021MNRAS.500.4042S, 2021AJ....161...51S}. From all types, exoplanet atmospheres free of clouds offer a unique opportunity to constrain elemental abundances (e.g., Na, K, water and carbon-bearing molecules and metallicity objectively), as clouds and hazes obscure the amplitudes of absorption features \citep{fortney05b, fortney10}. Such rare exoplanets enable unbiased absolute abundance constraints, which may inform theory of irradiated gas-giant exoplanet atmospheres and giant planet formation \citep{2011ApJ...743L..16O, 2012ApJ...758...36M, 2017MNRAS.469.4102M}. Because the composition of protoplanetary disks vary significantly with distance from the star, an exoplanet's atmospheric composition may contain a fossil record of the conditions of the disk at the location where the planet formed. \cite{2012ApJ...758...36M} showed that the C/O ratio carries important information on the chemistry of volatile species, which has started to provide definitve constraints \citep{2021Natur.598..580L}. \cite{2021ApJ...914...12L} explored the insight that the measurement of refractory abundances can provide into a planet’s origins. Through refractory-to-volatile elemental abundance ratios, they demonstrated that a planet’s atmospheric rock-to-ice fraction can be estimated, which can constrain planet formation and migration scenarios.

  %{\bf{You likely want to add a short paragraph on HJs citing relevant detections, clouds, ect, which will put Wasp-96 in perspective before talking specifically about it.}}

Motivated by the recent observational results from space and the ground, theoretical predictions for strong alkali features, stratosphere inducing TiO and VO, and diversity of clouds hazes and dust, we initiated a large transmission spectral survey of twenty exoplanets with the FORS spectrograph on the Very Large Telescope (VLT). Optical transmission spectra from this program have been reported for WASP-96b, WASP-103b, WASP-88b and WASP-110b \citep{2018Natur.557..526N, 2020MNRAS.497.5155W, 2021MNRAS.506.2853S, 2021AJ....162...88N}. In this study, we performed follow-up reconnaissance observations of WASP-96b using {\it{HST}} and {\it{Spitzer}}, aiming at detecting water in the transmission spectrum of the planet.

\begin{figure}
\begin{multicols}{1}
    \includegraphics[width=2\linewidth]{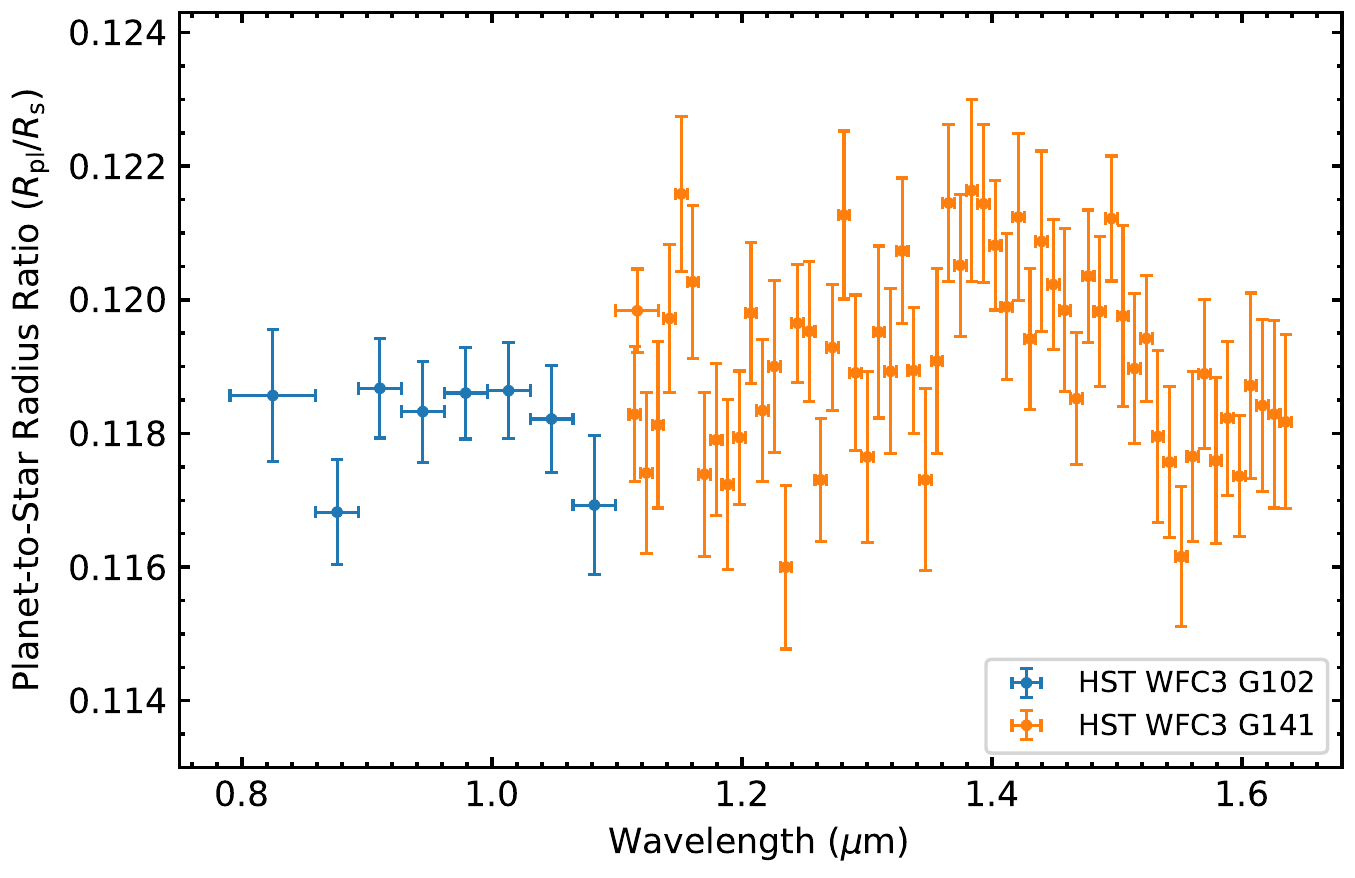}\par 
    \end{multicols}
\begin{multicols}{1}
    \includegraphics[width=2\linewidth]{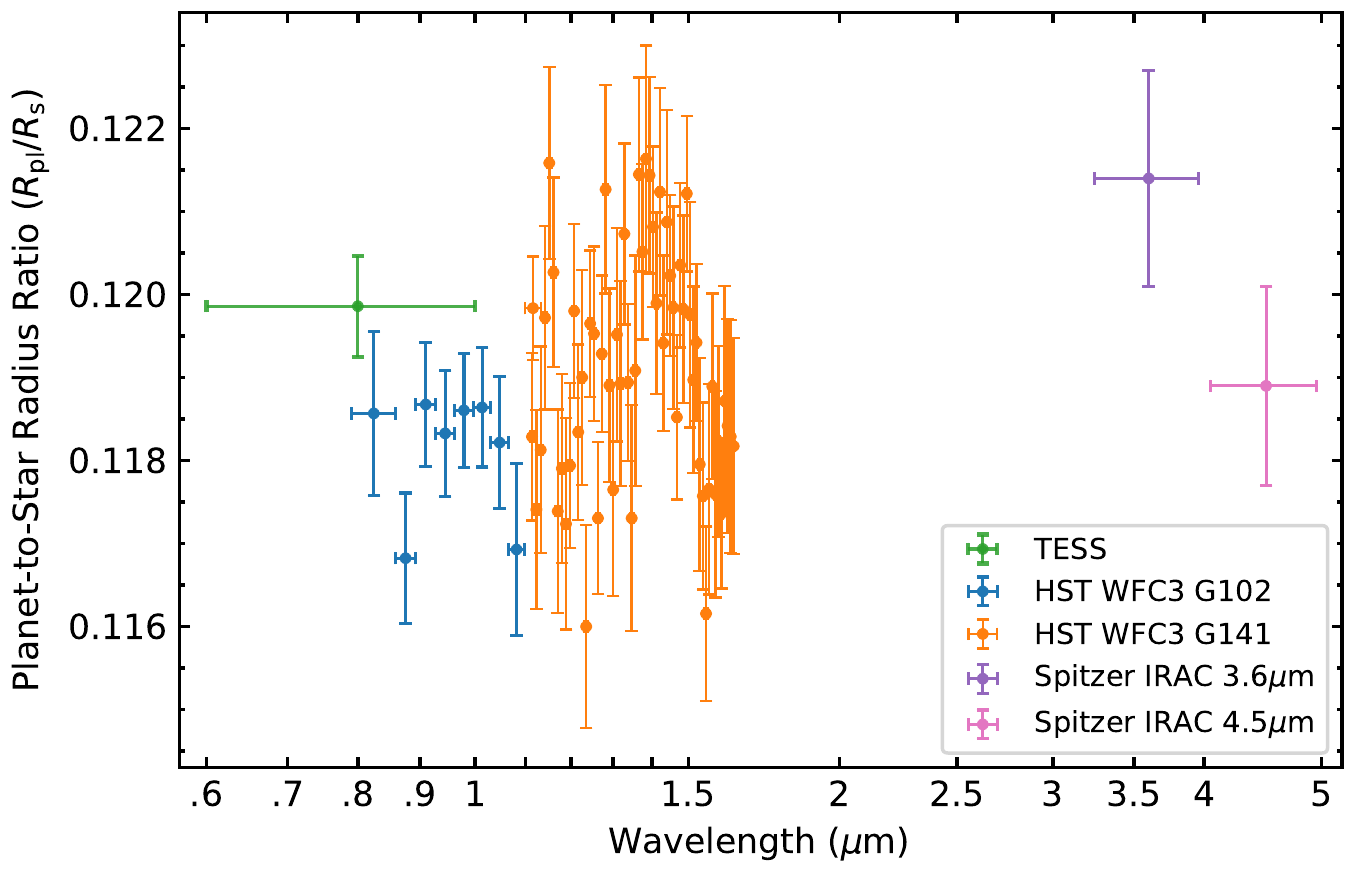}\par
\end{multicols}
\caption{Infrared transmission spectrum of the cloud-free WASP-96b. The error bars indicate the $1\sigma$ uncertainties. {\it{Top panel:}} Results from the two {\it{HST}} WFC3 spectroscopic observations. {\it{Lower panel:}} Combined  {\it{HST}} WFC3 transit spectroscopy and TESS and {\it{Spitzer}} IRAC $3.6\mu$m and $4.5\mu$m transit photometry.}
\label{fig:fig0A}
\end{figure}

% ($\log{\rm{R^{'}_{H\&K}}}=-5.3\pm0.1$ measured in this work)   and photometrically stable (r.m.s.$<0.1$mmag) 

{\mbox{WASP-96b}} is a ''hot Saturn`` with a planetary mass $M_{\rm{p}}=0.48\pm0.03M_{\rm{J}}$, where $M_{\rm{J}}$ is the mass of Jupiter, planet radius $R_{\rm{p}}=1.20\pm0.06R_{\rm{J}}$, where $R_{\rm{J}}$ is the radius of Jupiter and an equilibrium temperature \mbox{$T_{{\rm{eq}}}=1285\pm40$K} on a 3.4-day circular orbit around a ${\mathrm{V}}=12.2$ chromospherically quiet and photometrically stable (Section\,\ref{sec:tess_lcs}) G8 star at a distance of $356\pm5$\,pc in the southern constellation Phoenix \citep{hellier14}. Unlike the optical transmission spectrum for a majority of hot Jupiters known to date, the spectrum of WASP-96b shows very clear pressure-broadened wings of the sodium line, evidence of potassium and a near-UV Rayleigh scattering slope, with the latter defining the hydrogen continuum level and proving the planet has a clear atmosphere at the limb. The VLT spectrum provides an absolute sodium abundance, $2-18\times$ the solar value ($1\sigma$) and an atmospheric metallicity in agreement with the mass-metallicity trend observed for solar-system planets and exoplanets \citep{2018Natur.557..526N, 2016ApJ...831...64T}. This suggests that the absorption signature of molecules, such as water with absorption bands at $\sim0.95$, $1.15$ and $1.4\mu$m, carbon monoxide and methane at $3.6$ and $4.5\mu$m are expected to also be free from the suppressing effect of clouds and hazes in the infrared \citep{sing16}. We carried out (i) transit observations with the {\it{Hubble Space Telescope}} ({\it{HST}}) in spatial scanning mode, and (ii) {\it{Spitzer Space Telescope}} ({\it{Spitzer}}) transit and eclipse observations, to extend the planet's optical transmission spectrum at these wavelengths, constrain the atmospheric composition and day side brightness temperature of the planet \citep[GO-15469 and 14255;][]{2018hst..prop15469N, 2019sptz.prop14255N}. An analysis of the transit observations has been reported by \cite{2021AJ....161....4Y}, finding absorption from water and an offset between the optical and infrared. 

In this paper, we present new results for the infrared transmission spectrum, which overall agree with the results from \cite{2021AJ....161....4Y}, finding solar to supersolar abundances for sodium and oxygen. We additionally find the planet's potassium and carbon abundances to be low with broad uncertainties owing to lower precision of the spectrum at the relevant wavelengths. We present new results for the elemental abundances of the host star, its chromospheric activity index, constrain the dayside temperature of the planet and refine the system parameters and planet orbital ephemeris. We demonstrate in this work, as has been detailed in \cite{2018Natur.557..526N} that the observed offset of the optical spectrum is explained by the assumed absolute planet-to-star radius ratio for the calculation of common mode (wavelength independent) systematics model for the ground-based spectra. As the hot Jupiter with the clearest known atmosphere, WASP-96b is poised to become an important prototype exoplanet target for the {\it{James Webb Space Telescope}} and provide the first accurate abundance measurements, unbiased by clouds and hazes.

We describe the observations in Section\,\ref{sec:obs}, discuss the data reduction and light curve analysis in Section\,\ref{sec:lcan}, present results from a retrieval analysis in Section\,\ref{sec:res} and discuss our findings in the context of literature result in Section\,\ref{sec:disc} and summarize and conclude in Section\,\ref{sec:concl}.

%table of planet abundances along with stellar abundances in the same units

%This is a simple template for authors to write new MNRAS papers.
%See \texttt{mnras\_sample.tex} for a more complex example, and \texttt{mnras\_guide.tex}
%for a full user guide.

%All papers should start with an Introduction section, which sets the work
%in context, cites relevant earlier studies in the field by \citet{Fournier1901},
%and describes the problem the authors aim to solve \citep[e.g.][]{vanDijk1902}.
%Multiple citations can be joined in a simple way like \citet{deLaguarde1903, delaGuarde1904}.

\begin{figure}
\includegraphics[width=0.99\linewidth]{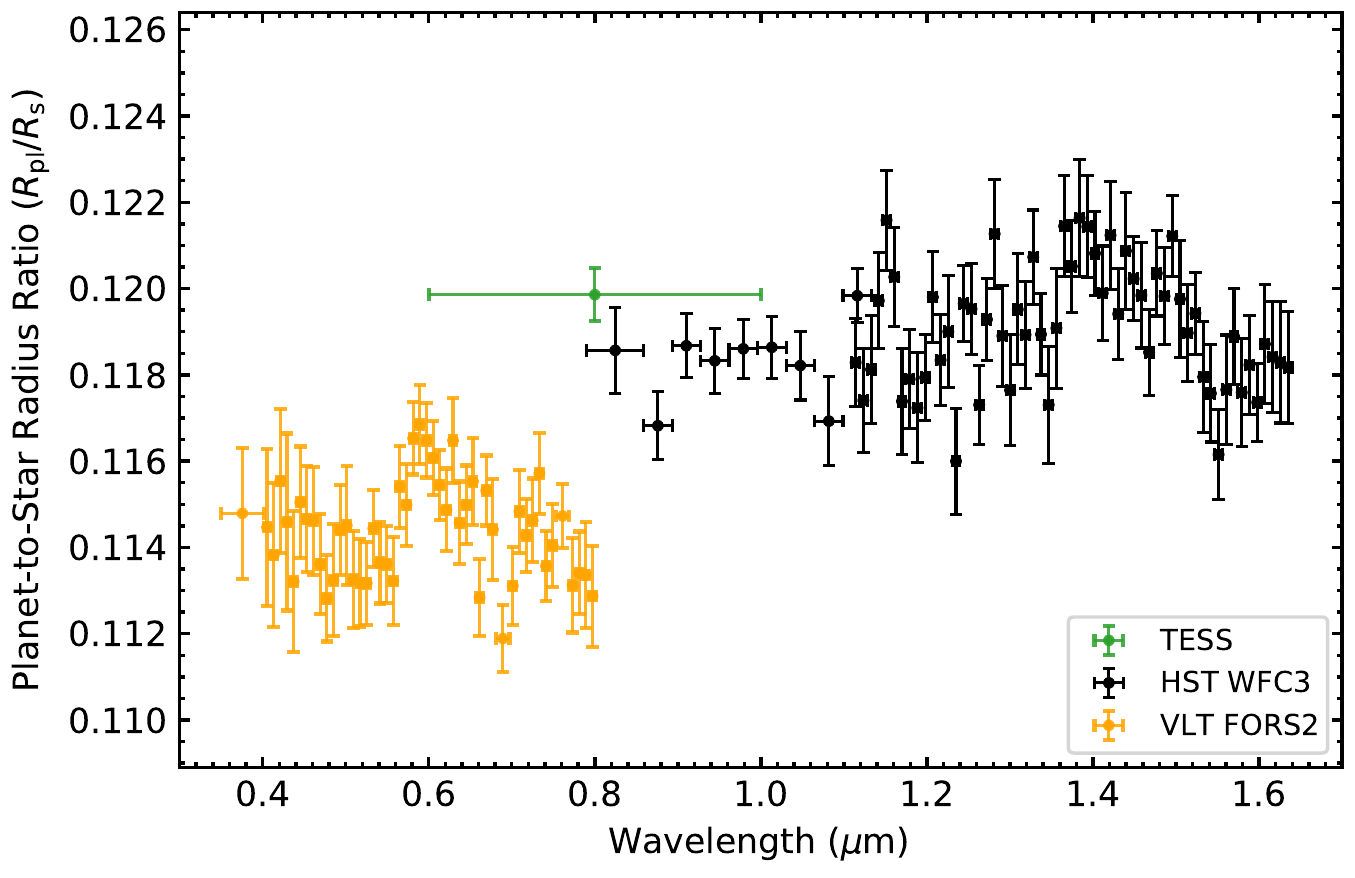}
\caption{Combined transmission spectrum of WASP-96b. Plotted are the VLT and {\it{HST}} spectra with orange and black symbols, respectively. A radius measurement from a combined analysis of TESS photometric light curves is indicated with the green symbol. Error-bars mark the $1\sigma$ uncertainties.}
\label{fig:fig0B}
\end{figure}

\section{Observations and reductions}\label{sec:obs}
\subsection{HST WFC3 spatial scan mode transit spectroscopy}\label{sec:hst_obs}

We observed two primary transits of WASP-96b with the Wide Field Camera 3 (WFC3) aboard the {\it{Hubble Space Telescope}} ({\it{HST}}) as part of Program 15469, \citep{2018hst..prop15469N} in spatial scan mode. Both visits included five consecutive {\it{HST}} orbits, with the first, second, third and fifth orbits covering out of transit phases and the fourth orbit during the transit. The {\it{HST}} orbital sequence was planned such that the in-transit orbit covers the planet phase between second and third contacts (end of ingress and start of egress, respectively) to maximize the precision of the measured planet radius (Figure\,\ref{fig:figA1} and\,\ref{fig:figA2}). Each observation was performed in spatial scan mode using the $256\times256$ subarray centered on the target to reduce read-out overheads. The spatial scan mode spreads the stellar flux across more pixels compared to a fixed pointing (stared) observation, allowing for a longer exposure times with higher duty cycle. Prior to the spectral observations, we acquired direct images of the target to facilitate a wavelength solution and to inspect the vicinity of the target for field stars that can contaminate the spectra. We exposed for 1.111 and 1.389s using the F127M and F139M filters for the G102 and G141 grisms, respectively. An inspection of the direct images showed the field around WASP-96 to be free of stars.

The first observation covers the transit on UT 2018 December 18th. We used the dispersive element G102, which covers the spectral range from 0.8 to $1.15\mu$m at a resolving power of $R\sim210$ (at $1\mu$m) and a dispersion of $2.45$\,nm\,px$^{-1}$. We used forward scanning with a rate of 0.013 arcsec sec$^{-1}$ and the SPARS25 sampling sequence with eleven iterations per exposure (NSAMP9) resulting in total integration times of $\sim179$\,sec and scans across 2.3 arcsec $(\sim18$\,px$)$. Typical count levels reached a maximum of $\sim2.9\times10^4$ analogue-to-digital units (ADU), i.e. well within the linear regime of the detector. 

The second observation covered the transit on UT 2018 December 28th. We exploited the dispersive element G141, which covers the spectral range from 1.075 to $1.7\mu$m at a resolving power of $R\sim130$ (at $1.4\mu$m) and a dispersion of $4.65$\,nm\,px$^{-1}$. We used forward scanning with a rate of 0.022 arcsec sec$^{-1}$ and the SPARS25 sampling sequence with twelve iterations per exposure (NSAMP8) resulting in total integration times of $\sim157$\,sec and scans across 3.4 arcsec $(\sim268$\,px$)$. The spectra reached typical maximum count levels of $\sim2.8\times10^4$ ADU, i.e. within the linear regime of the detector.

We performed data reduction and analysis following the procedures outlined in \cite{nikolov18a}. Our analysis started with the 2D  {\texttt{ima}} spectra, produced by the {\texttt{calwf3}} pipeline (v3.1), which are corrected for bias, dark current, flat-field and detector non-linearity effects. We extracted flux for WASP-96b from each exposure by taking each successive non-destructive read difference. We removed the background on each individual read difference by taking the median flux in a box away from the stellar spectrum. We then determined the flux weighted centre of the WASP-96 scan and set to zero all pixel values located more than 20 pixels above and below along the cross-dispersion axis. Application of this top hat filter had the effect of eliminating most of the pixels affected by cosmic rays. The final reconstructed images were produced by adding together the read differences for each exposure. Any remaining cosmic rays were identified and corrected, following the procedure described in \citet{nikolov14, nikolov18a}. We find a total of 2 or 3 pixels that sample the spectra to be affected by cosmic rays for each reconstructed image.

We performed a spectral extraction using a fixed-size box by summing the flux of all pixels. The box had dimensions of $171\times28$ and $169\times39$ pixels for the G102 and G141 data, respectively and centered for each individual exposure. To identify the box positions along the dispersion and cross-dispersion axis we took the flux-weighted mean of each 2D spectrum. The target drift along the dispersion and cross-dispersion axis are shown in Figure\,\ref{fig:figA1}. 

We established wavelength solutions for each grism time series by cross-correlating each spectrum with an ATLAS synthetic stellar spectrum \citep{kurucz79} that matches the astrophysical properties of the WASP-96 host star i.e., ${\mathrm{T_{\rm{eff}}=5500\pm150}}$\,K, $\log{g}=4.42\pm0.02$ and ${\mathrm{[Fe/H]}} =0.14\pm0.19$, reported in \citet{hellier14}. We obtained synthetic spectra from the {\it{HST}} exposure time calculator, which also accounts for the instrument sensitivity.

\subsection{Spitzer IRAC transit and eclipse photometry}\label{sec:spitzer_obs}
Photometric data were collected during two primary transits and two secondary eclipses using the {\it{Spitzer Space Telescope}} ($Spitzer$, \citealt{Werner04})  Infrared Array Camera (IRAC, \citealt{Fazio04}), as part of Program 14255, \cite{2019sptz.prop14255N}. Each IRAC observation covers equal duration of in-transit,  pre-ingress and post-egress time series (Figure\,\ref{fig:figA3a} and \ref{fig:figA3b}). Each exposure consisted of 64 subarray frames of $32\times32$ pixels. We employed an exposure time of 1.92s and obtained a total of 9600 images for each observation. Details of the IRAC observations are summarized in Table\,\ref{tab:obstab}.

\begin{table}
\centering
\caption{{\it Spitzer} IRAC observations}
\label{tab:obstab}
\begin{tabular}{cccc} % four columns, alignment for each
\hline
\hline
\vspace{-0.05cm}
Channel & Start UTC & End UTC &  Event \\
\hline
ch1/3.6 & 2019-10-22\,10:29:59 & 2019-10-22\,15:51:20 & transit \\
ch2/4.5 & 2019-10-29\,06:59:07 & 2019-10-29\,12:20:28 & transit \\
ch1/3.6 & 2019-10-30\,23:54:54 & 2019-10-31\,05:16:15 & eclipse \\
ch2/4.5 & 2019-11-06\,20:15:01 & 2019-11-07\,01:36:22 & eclipse \\
\hline
\end{tabular}
\end{table}

Our data reduction and analysis procedures follow the methodology detailed in \citet{nikolov15}. The data were calibrated by the {\it{Spitzer}} pipeline version 21 and are available in the form of Basic Calibrated Data {\texttt{(.bcd)}} files. After organizing the data, we converted the images from flux in mega-Jansky per steradian (MJy sr$^{-1}$) to photon counts i.e., electrons using the information provided in the {\sc{FITS}} headers. We then performed an outlier filtering for hot (energetic) or lower pixels in the data by following each pixel through time. This task was performed in two steps, first flagging all pixels with intensity more than $8\sigma$ compared to the median value computed from the five preceding and five following images. The values of these flagged pixels were replaced with the local median value. In the second pass, we flagged and replaced outliers above the $4\sigma$ level, following the same procedure. The total fraction of corrected pixels was $\sim0.4$ per cent for the 3.6 $\mu$m and $\sim0.1$ per cent for the 4.5 $\mu$m channel.

We then subtracted the background flux from the time series by performing an iterative $3\sigma$ outlier clipping. For each image we removed the pixels within the stellar point spread function (PSF), background stars or hot pixels. We then created a histogram from the remaining pixels and fitted a Gaussian to determine the sky background. Prior to aperture photometry, we measured the target's PSF position on each image using the flux-weighted centroiding method with a circular region with a radius of 3 pixels. The variation of the $x$ and $y$ positions of the PSF on the detectors were measured to be $\sim0.2$ for the 3.6 $\mu$m and $\sim0.1$ pixels for the 4.5 $\mu$m transits. We measure variation of the $x$ and $y$ positions of the PSF of $\sim0.5$ and $\sim0.1$ for the 3.6 $\mu$m and $\sim0.3$ and $\sim0.4$ for the 4.5 $\mu$m eclipses. 

We performed photometry with both fixed and time variable apertures and filtered the resulting light curves for $5\sigma$ outliers with a width of 20 data points. In the first method we covered the range of fixed (in time) radii from 1.5 to 3.5 pixels in increments of 0.1 pixels. In the second approach, we scaled the size of the extraction aperture by the value of a quantity known as the noise pixel parameter \citep{mighell05, knutson12}. The best results from both methods were identified by examining both the residual root mean square (rms) after fitting the light curves from each channel, as well as the white and red noise components measured with the wavelet technique detailed in \cite{carter09}. While we find no evidence for a particular aperture size that produces significantly lower light curve scatter compared to all apertures, the fitted transit depths are consistent within their $1\sigma$ uncertainties. The time variable approach produced the lowest light curve scatter for the $3.6\mu$m observations with an aperture radii of 1.97 and 2.05 pixels with sky annuli from 2.10 to 5.72 and from 2.05 to 5.54 for the transit and eclipse, respectively. The lowest light curve scatter for the $4.5\mu$m is found from the fixed and time variable aperture methods for the transit and eclipse, respectively. We find aperture radii of 2.25 and 2.43 pixels and sky annuli defined between radii 2.25 and 6.08 pixels and from 2.43 to 6.55 pixels for the transit and eclipse, respectively (Figure\,\ref{fig:figA3a} and \,\ref{fig:figA3b}).

\subsection{TESS transit photometry}\label{sec:tess_obs}
WASP-96 was observed by the Transiting Exoplanet Survey Satellite \citep[TESS]{2014SPIE.9143E..20R} on Camera 2 CCD 1 during the Prime and Extended Mission between 2018 August 23 (Sector 2) and September 20, and 2020 August 26 and September 21 (Sector 29) with a 2-minute cadence in the full frame images (FFIs). We obtained light curves for both observing campaigns from the MAST archive. The light curves were produced by the TESS Science Processing Operations Center (SPOC) pipeline, which was used to calibrate full-frame images (FFI) and to assign world-coordinate system information to the FFI data delivered to the MAST \citep{2020RNAAS...4..201C}. First we inspected the simple aperture photometry (SAP) fluxes and identified seven transits in each campaign. We chose to perform light curve fits with the goal to identify and remove residual systematic effects. We obtained portions of the light curves centered on the primary transits with six hour out-of-transit baseline data prior ingress and additional six hours after egress. One transit from the second campaign has been discarded in our analysis due to a significantly higher flux scatter as compared to the rest of the data.

\begin{figure*}
\centering
\includegraphics[trim={0.0cm 0.0cm 0.0cm 0.0cm}, width=0.9\linewidth]{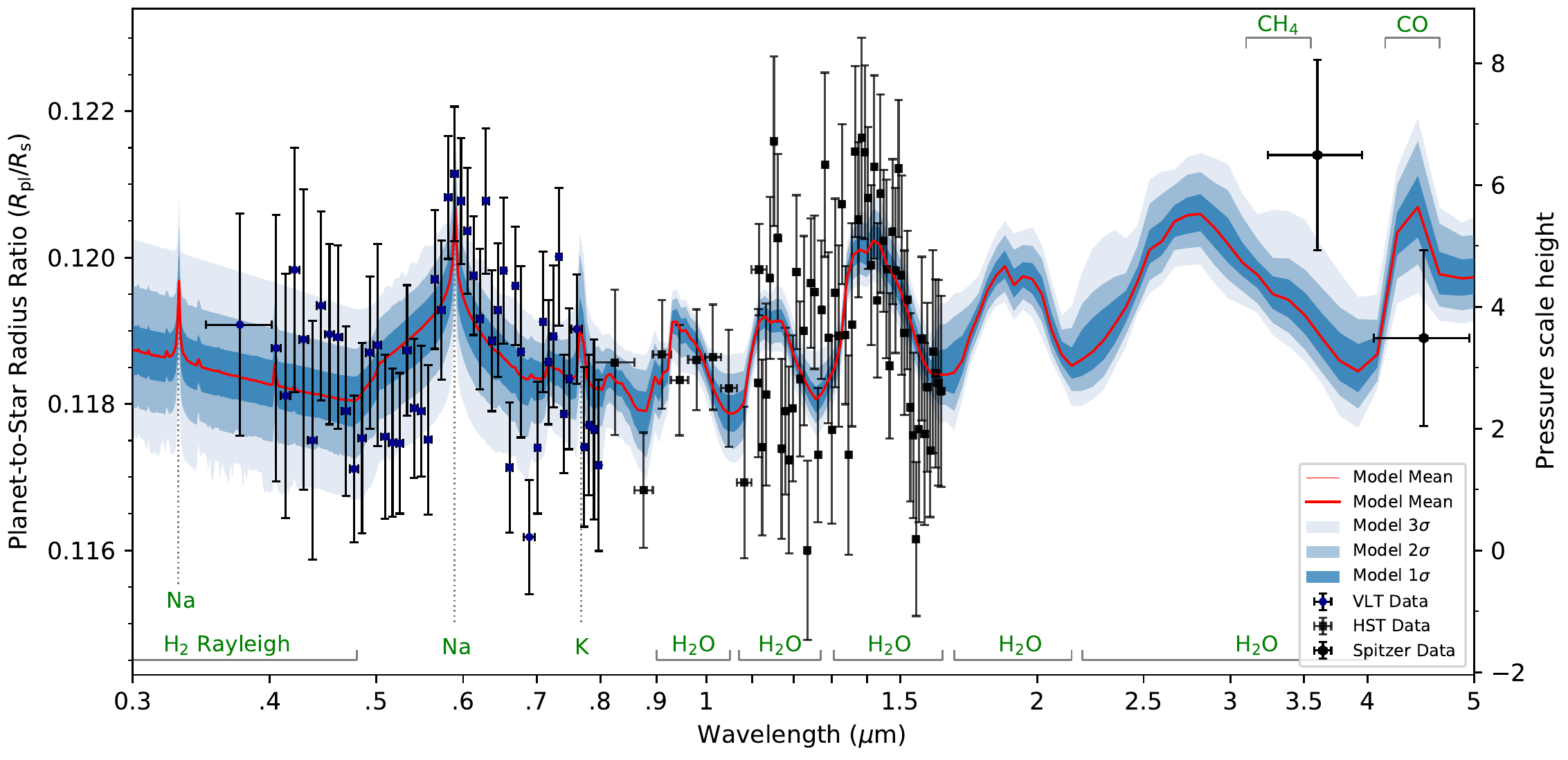}
\caption{Comparison of the VLT, {\it{HST}} and {\it{Spitzer}} measured transmission spectra of WASP-96b (dots with $1\sigma$ vertical error bars; the bin widths are indicated with horizontal bars) with the best-fit model from the retrieval analysis (red line) with resolution identical to the data, with the $1\sigma$, $2\sigma$ and $3\sigma$ confidence levels (dark to pale blue regions).}
\label{fig:fig1}
\end{figure*}

\begin{table}
\centering
\caption{System parameters from {\it{HST}} WFC3}
\label{tab:systpartab}
\begin{tabular}{lc} % four columns, alignment for each
\hline
\hline
\vspace{-0.05cm}
 Parameter, symbol, unit & Value \\
\hline
Orbital Period, $P$, (day) & 3.4252602 (fixed) \\
eccentricity, $e$ & 0 (fixed)\\
inclination, $i, (^{\circ})$ & $85.14$ (fixed)\\
$a/R_{\ast}$ & $8.84$ (fixed)\\
\hline
$HST\,G102$ & $0.8-1.15\mu$m\\
\hline
Mid-transit time, ${\rm{T_{mid}}}$, JD	& $2458470.7795^{+0.0037}_{-0.0015}$\\
Planet-to-star radius ratio, $\mathrm{R_{\rm{p}}/R_{\ast}}$ & $0.1186^{+0.0015}_{-0.0013}$\\
First limb darkening coefficient, $c_1$ & $0.631621$ (fixed)\\
Second limb darkening coefficient, $c_2$ & $-0.251583$ (fixed)\\
Third limb darkening coefficient, $c_3$  &	$0.417851$ (fixed)\\
Fourth limb darkening coefficient, $c_4$  & $-0.190682$ (fixed)\\
Characteristic amplitude, $A$ (ppm) & $2226^{+5369}_{-991}$\\
Characteristic length scale, $\eta_{\phi}$ & $4\pm{3}$\\
Characteristic length scale, $\eta_x$  & $7^{+4}_{-5}$\\
Characteristic length scale, $\eta_y$  & $2^{+5}_{-2}$\\
Intercept, $a_0$ & $0.9995^{+0.0016}_{-0.0018}$\\
Slope, $a_1$ & $0.00049^{+0.00048}_{-0.00034}$\\
Re-scaled uncertainty, $\sigma_w$, (ppm) & $158^{+78}_{-64}$\\
\hline
$HST\,G141$ & $1.075-1.700\mu$m\\
\hline
Mid-transit time,${\rm{T_{mid}}}$ JD	& $2458481.05626^{+ 0.00041}_{-0.00046}$\\
Planet-to-star radius ratio, $R_{\rm{p}}/R_{\ast}$ & $0.11962^{+0.00029}_{-0.00037}$\\
First limb darkening coefficient, $c_1$ & $0.657415 $ (fixed)\\
Second limb darkening coefficient, $c_2$ & $-0.105797 $ (fixed)\\
Third limb darkening coefficient, $c_3$  & $0.031906 $ (fixed)\\
Fourth limb darkening coefficient, $c_4$  & $-0.019303 $ (fixed)\\
Characteristic amplitude, $A$ (ppm), $A$ (ppm)	 & $2038^{+3167}_{-1156}$\\
Characteristic length scale, $\eta_{\phi}$  & $4\pm2$  \\
Characteristic length scale, $\eta_x$ & $8\pm4$\\
Characteristic length scale, $\eta_y$  & $4\pm2$\\
Intercept, $a_0$ & $0.9994\pm0.002$\\
Slope, $a_1$ & $0.00012^{+0.00028}_{-0.00025}$\\
Re-scaled uncertainty, $\sigma_w$, (ppm) & $128\pm24$\\
\hline
\end{tabular}
\end{table}

\subsection{ASSASN phase curve photometry}
WASP-96 was monitored for photometric variability by the Ohio State University’s All-sky Automated Survey for Supernovae (ASAS-SN) Photometry Database, \citet{Shappee2014, Jayasinghe2019}. A total of 233 photometric measurements were collected and processed during five consecutive observing campaigns starting on May 12, 2014 and ending on September 24, 2018. Each campaign, except the last one, covers the full visibility window of WASP-96 from May to January.

\subsection{ESO/MPG 2.2 FEROS high resolution spectroscopy}
The host star of WASP-96b was observed with the MPG/ESO 2.2-metre telescope equipped with the FEROS high-resolution spectrograph (ESO program: 098.A-9007, PI P. Sarkis). Two observations with an exposure time of 900 sec were obtained on UT 2016 December 12th and 19th, using a fiber with a diameter of 2\,arcsec and a spectral resolution of $R=\lambda/\delta\lambda=48,000$. The spectra were reduced with ESO's FEROS pipeline to phase three products i.e., extracted one-dimensional spectra with wavelength solution. Both spectra have a signal-to-noise ratio of 27 over a wavelength bin of $0.03\AA$. A magnified view centered on the  Ca\,{\sc{II}}\,H\&K lines from the combined spectra is shown in Figure\,\ref{fig:figHK}.

\section{Light curve analysis}\label{sec:lcan}
Our light curve analysis of the {\it{HST}}, {\it{Spitzer}} and TESS data follows the methods detailed in \citet{nikolov15, nikolov18a, 2021AJ....162...88N}.

\subsection{HST WFC3 spectrophotometry} \label{sec:wfc3_lcan}
\subsubsection{White-light curves}\label{sec:wfc3_wlc}

We produced white-light curves by summing the flux of each spectrum from 0.78 to 1.13 and from 1.12 to 1.65$\mu$m for the G102 and G141 transit time series, respectively (Figure\,\ref{fig:figA2}). Both data sets exhibit the well-known {\it{``hook''}} systematic, that correlates with the {\it{HST}} orbital phase and is considered to originate from charge-trapping in the WFC3 detector \citep{deming13, huitson13, zhou17}. The out-of-transit flux also exhibits a longer term drift, which is approximately linear in time.

We fit the transit and systematic effects of the white-light curves by treating the data as a Gaussian process\footnote{We made use of the publicly available {\tt{george}} Gaussian Process Python suite \citep{Foreman-Mackey15}.} \citep{gibson12a}. The transit parameters: orbital period, eccentricity, inclination $i$ and normalized semi-major axis $a/R_{\ast}$ (where $R_{\ast}$ is the radius of the star) were held fixed to the previously determined values of \citet{Anderson2014} and \citet{2018Natur.557..526N}, which are detailed in Table\,\ref{tab:systpartab}. We accounted for the stellar limb darkening by adopting the four-parameter non-linear limb-darkening law \citep{claret00} and computed the values of the coefficients ($c_1$, $c_2$, $c_3$, $c_4$) using a three-dimensional stellar atmosphere grid \citep{magic2015}. In these calculations, we adopted the closest match to the effective temperature, surface gravity and metallicity of the exoplanet host star found in \citet{hellier14}. The mid-time $T_{\rm{mid}}$ and planet-to-star radius ratio ${\mathrm{R_{p}/R_{\ast}}}$ were allowed to vary in the fit to each of the two white-light curves.

\begin{figure*}
\includegraphics[width=0.9\linewidth]{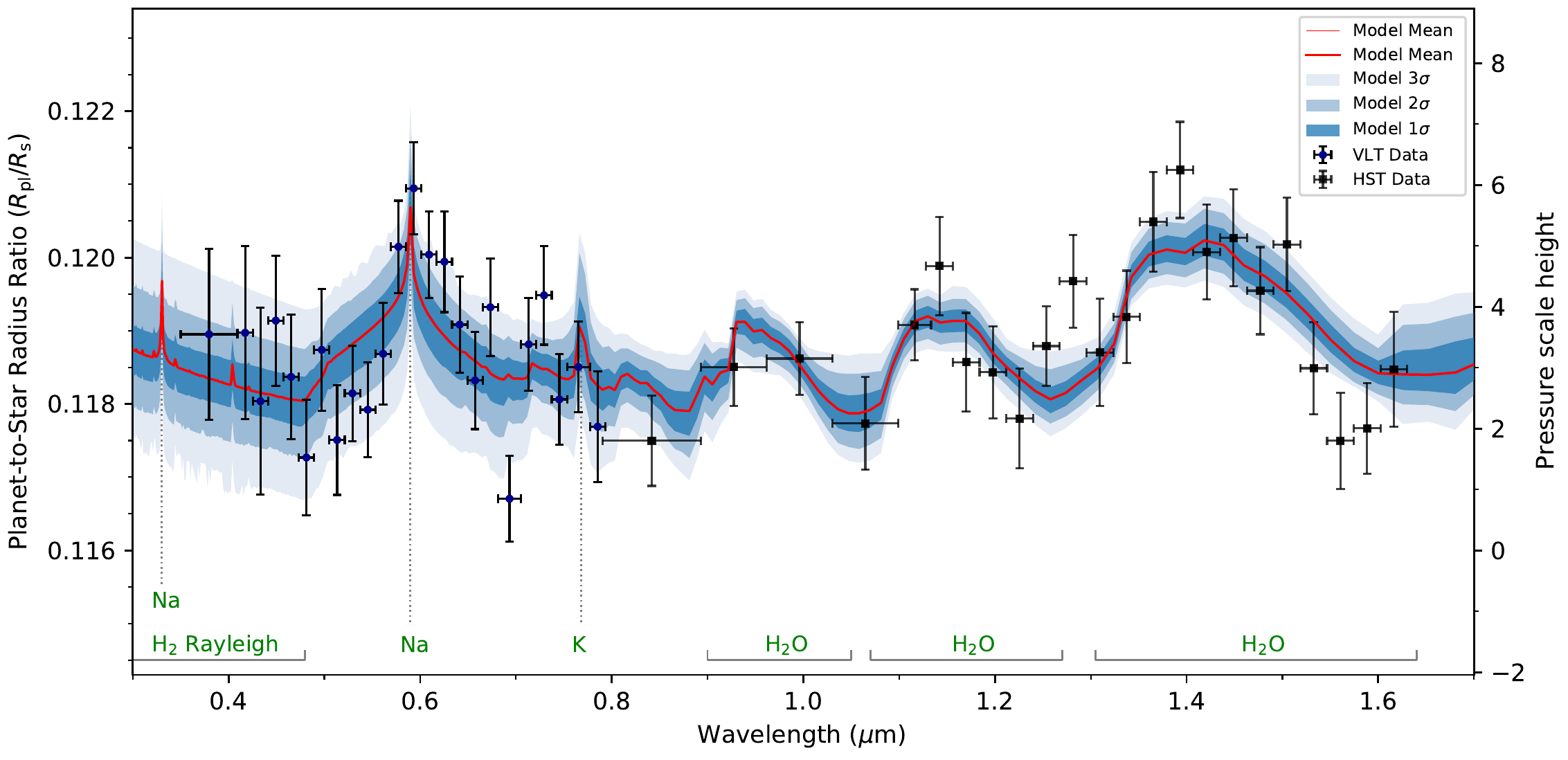}
\caption{Same as Figure\,\ref{fig:fig1} but showing a zoom around the combined optical-infrared spectrum from VLT and {\it{HST}}. We binned with the weighted mean the VLT, G102 and G141 spectra in 2, 2 and 4 points, respectively. }
\label{fig:fig2}
\end{figure*}

Under the Gaussian process assumption, the data likelihood is a multivariate normal distribution with a mean function $\mu$ describing the deterministic transit signal and a covariance matrix $K$ that accounts for stochastic correlations (i.e., poorly constrained systematics) in the light curves:
\begin{equation}
p(\bm{f} \vert \bm{\theta}, \gamma) = \mathcal{N}(\mu,K),
\end{equation}

where $p$ is the probability density function, $\bm{f}$ and $\bm{\theta}$ are vectors containing the flux measurements and mean function parameters, respectively; $\gamma$ is a function containing the covariance parameters and $\mathcal{N}$ is a multivariate normal distribution. The mean function $\mu$ is defined as:
\begin{equation}
\mu(\bm{t},\bm{\hat{t}};a_0,a_1,\bm{\theta})=[a_0+a_1\bm{\hat{t}}]\,T(\bm{\emph{t}},\bm{\theta}),
\end{equation}

where $\bm{t}$ is a vector of all central exposure time stamps in Julian Date, $\bm{\hat{t}}$ is a vector containing all standardized times, that is, with subtracted mean exposure time and divided by the standard deviation, $a_0$ and $a_1$ describe a linear baseline trend, $T(\bm{\theta})$ is an analytical expression describing the transit and $\bm{\theta}=(i, a/R_{\ast}, T_{\rm{mid}}, R_{p}/R_{\ast}, c_1, c_2, c_3, c_4)$. We made use of the analytical formulae of \citet{mandel02, kreidberg2015}.

Similar to our earlier {\it{HST}} and VLT studies, we defined the covariance matrix as $K = \sigma_i^2\delta_{ij} + k_{ij}$, where $\sigma_i$ contains the photon noise uncertainties, $\delta_{ij}$ is the Kronecker delta function and $k_{ij}$ is a covariance function. The white noise term $\sigma_w$ was assumed to have the same value for all data points and was allowed to freely vary. We chose to use the Mat\'{e}rn $\nu = 3/2$ kernel with the {\it{HST}} orbital phase ($\phi$) and spectral dispersion $(x)$ and cross-dispersion $(y)$ drifts, respectively as input variables. Our kernel choice is motivated by the study of \cite{gibson13b}, where the Mat\'{e}rn $\nu = 3/2$ kernel is empirically motivated using simulated data, and is the first to use this kernel for light curve analysis. The GP free parameters were the characteristic correlation amplitude $(A)$ and correlation length scales for each input parameter $(\tau_{\phi}, \tau_{x}, \tau_{y})$. Similar to our previous studies, we also accounted for the long-term out-of-transit baseline trend with a linear time term. As with the linear time term, we also standardized the input parameters before the light curve fitting. The covariance function was defined as:
\begin{equation}
k_{ij} = A^2(1+\sqrt{3}D_{ij})\exp{-\sqrt{3}D_{ij}},
\end{equation}

where $A$ is the characteristic correlation amplitude and

\begin{equation}
D_{ij} = \sqrt{\frac{(\bm{\hat{\phi}}_i-\bm{\hat{\phi}}_j)^2}{\tau_\phi^2}+\frac{(\bm{\hat{x}}_i-\bm{\hat{x}}_j)^2}{\tau_x^2}+\frac{(\bm{\hat{y}}_i-\bm{\hat{y}}_j)^2}{\tau_y^2}},
\end{equation}

where $\tau_{\phi}, \tau_{x}, \tau_{y}$ are the correlation length scale and the hatted variables are standardized. The parameters $\bm{X} = (a_0, a_1, T_{\rm{mid}}, R_{p}/R_{\ast}, \sigma_w)$ and $\bm{Y} = (A, \tau_\phi, \tau_x, \tau_y)$ were allowed to vary and fixed the orbital period $P$ and eccentricity $e$ to the values reported in \cite{hellier14} and the system parameters $(i)$ and $(a/R_{\ast})$ to the values found in \citet{2018Natur.557..526N}.Our choice to fix the two system parameters was determined by the fact that both {\it{HST}} observations lack data points (except one for the last orbit of the G141 visit, Figure\,\ref{fig:figA2}) obtained during ingress and egress. We adopted uniform priors for $x$ and log-uniform priors for $y$. We marginalize the posterior distribution by making use of the Markov-Chain Monte Carlo (MCMC) software package {\tt{emcee}} \citep{Foreman13}. We identified the maximum likelihood solution using the Levenberg–Marquardt least-squares algorithm \citep[][]{markwardt09} and initialized three groups of $250$ walkers close to that maximum. The first two groups were run for $450$ samples and the third one for $3,500$ samples. To ensure faster convergence, we re-sampled the positions of the walkers in a narrow space around the position of the best walker from the first run before running for the second group. This helps prevent some of the walkers starting in a low-likelihood area of parameter space, which can require more computational time to converge. Figure\,\ref{fig:figA2} and Table\,\ref{tab:systpartab} show transit models for each of the two observations computed using the marginalized posterior distributions and the fitted parameters, respectively. We find residual dispersion of 83 and 94 parts-per-million for the blue (G102) and red (G141) light curves, respectively.

\begin{figure}
\centering
\includegraphics[trim={1.2cm 0.0cm 0cm 0cm}, width=0.98\linewidth]{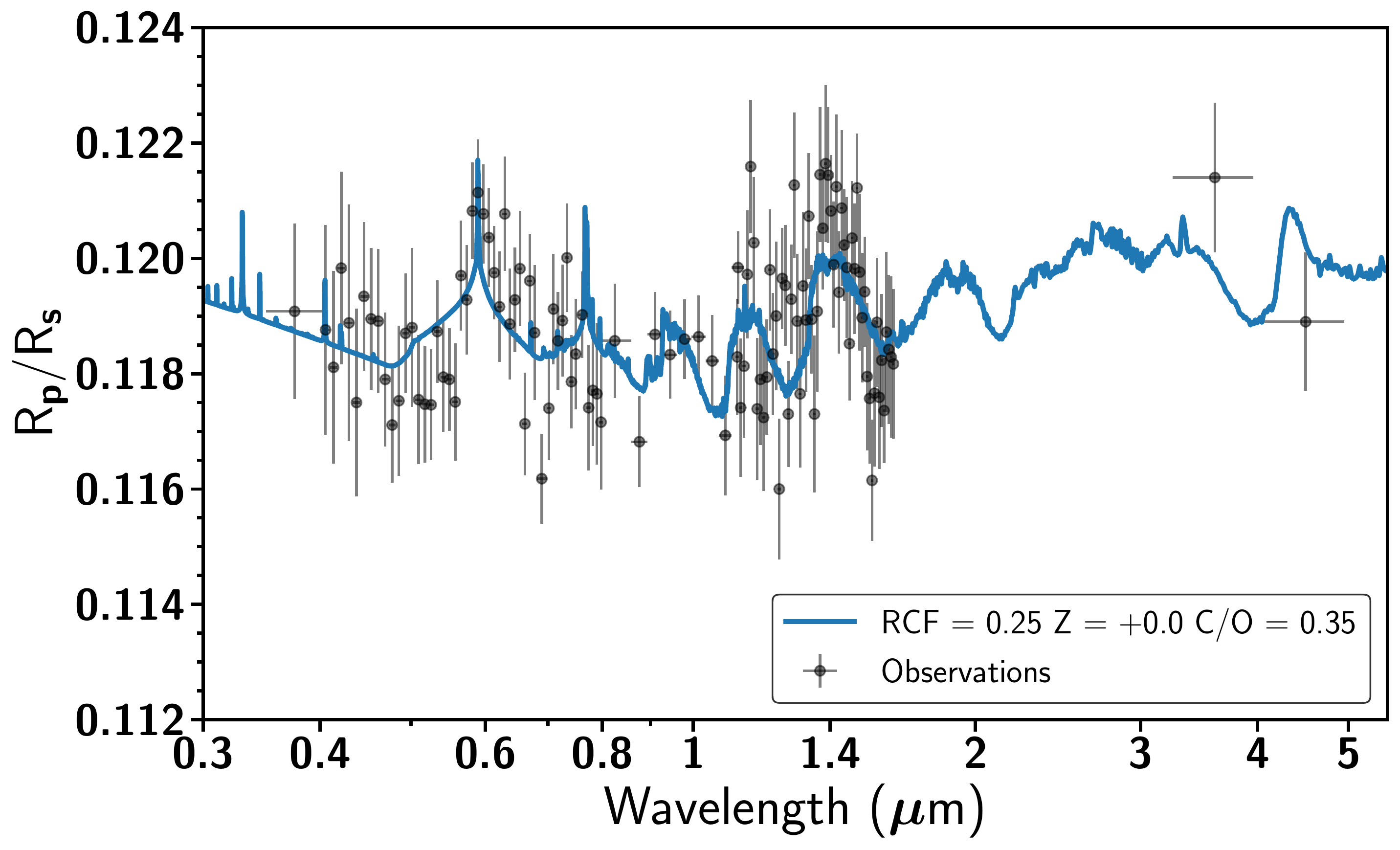}
\caption{Comparison of the VLT, {\it{HST}} and {\it{Spitzer}} measured transmission spectrum (black dots with 1$\sigma$ error bars) with the best fit forward model transmission spectrum (blue) from the self-consistent planet-specific grid of WASP-96b across a range of re-circulation factor (RCF), metallicity (Z) and C/O ratio.  $\chi^2 = 129.3$ for the best fit model spectrum. The best fit model parameters are RCF = 0.25, solar metallicity and C/O ratio of 0.35.}
\label{fig:forwd}
\end{figure}

\begin{figure*}
\includegraphics[width=0.80\linewidth]{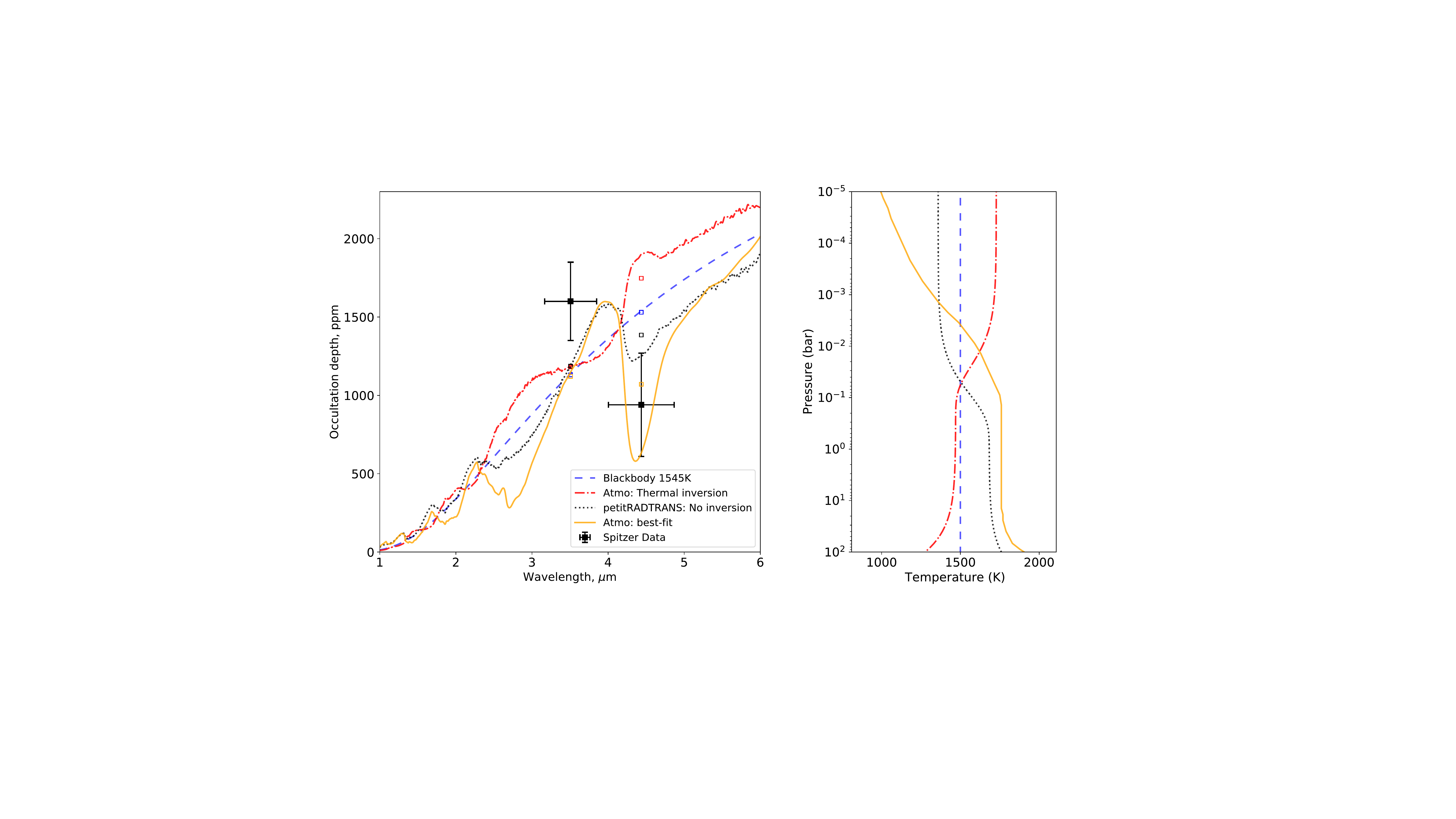}
\caption{Emission spectrum of WASP-96b, synthetic emission spectra and assumed pressure-temperature profiles. {\it{Left panel:}} WASP-96b thermal spectrum from {\it{Spitzer}} IRAC 3.6 and 4.5 $\mu$m observations (black symbols with $1\sigma$ error bars) compared to forward synthetic spectra assuming an isothermal (dashed line), {\tt{ATMO}} and {\tt{petitRADTRANS}} decreasing (continuous and dotted lines, respectively) and inverted (dash-dotted line) pressure-temperature profiles. Model predictions, averaged over the wavelengths for each channel are indicated with the open symbols and color coded to the relevant model. The lowest chi-square spectrum corresponds to the {\tt{ATMO}} synthetic spectrum, assuming a non-inverted pressure-temperature profile (continuous line). Given the relatively low level of irradiation for WASP-96b, an inverted temperature profile is rather unlikely. Yet, we include a spectrum for the inverted P-T profile merely for reference purposes, not expecting it to fit the data. {\it{Right panel:}} Pressure-temperature profiles assumed in the calculation of the thermal emission spectra. }
\label{fig:fig4}
\end{figure*}

\subsubsection{Spectroscopic light curve analyses}\label{sec:wfc3speclc}
We produced spectroscopic light curves by summing the flux of WASP-96 in bands with variable widths in the range from 0.7903 to 1.6396$\mu$m. Our choice for the bin widths was governed by the goal to obtain the highest spectral resolution with spectrophotometric precision that is comparable to the precision of the VLT data. We produced 9 and 56 bins for the G102 and G141 grisms, respectively. The G102 and G141 grisms overlap in wavelength over a narrow region from 1.0195 to 1.1333$\mu$m, which allows a direct comparison of the measured transit depths, Figure\,\ref{fig:fig1} and \ref{fig:fig2}, Table\,\ref{tab:transpec}. 

We established wavelength-independent i.e., common-mode, systematic correction factors for each data set, similar to our VLT and {\it{HST}} studies detailed in \cite{nikolov14, nikolov16}. We obtained the correction factors by dividing the white-light curve from each visit by the transit model computed with the parameters detailed in Table\,\ref{tab:systpartab}. The common-mode factors for each visit are shown in Figure\,\ref{fig:figA2}.

We fit the spectroscopic light curves using a two-component function that takes into account the transit signal and systematics simultaneously, similar to our studies in \citet{nikolov16, 2018Natur.557..526N}. Prior to fitting for the systematics, we removed the common mode factors from each set of spectroscopic light curves by dividing the raw flux of the spectroscopic light curves by the corresponding common-mode factors for the relevant visit. We computed transit models using the analytical models of \citet{mandel02}. In the fits, we allowed only the planet-to-star radius ratio $R_{p}/R_{\ast}$ to vary. We held the four coefficients of the nonlinear limb darkening law to their theoretical values. We obtained these values following the same method as for the white light curves.

We accounted for the systematics using a fourth-order polynomial of {\it{HST}} orbital phase ($\phi$) and up to a second order-polynomial of dispersion and cross-dispersion drift, Figure\,\ref{fig:figA1}. We produced all possible combinations of systematics models and performed separate fits with each of them included in the two-component function. This approach has been preferred as opposed to GP regression, as the CM-corrected {\it{HST}} spectroscopic light curves exhibit a lower level of systematic effects \citep{sing15, sing16, nikolov18a}. We computed the Akaike Information Criterion (AIC) for each fit and estimated the statistical weight of the model depending on the number of degrees of freedom, (\citealt{akaike74}). We marginalized the resulting relative radii following \cite{gibson14}. The highest evidence models included a fourth order polynomial of {\it{HST}} orbital phase.

%Prior to each light curve fit, we set the spectrophotometric uncertainties of each band to values with additional component from readout noise. The best-fit models were identified using a Levenberg-Marquardt least-squares algorithm and rescaled the uncertainties of the fitted parameters using the dispersion of the residuals, \citep[][]{markwardt09}. All residual outliers larger than $3\sigma$, typically 1-2 only in a few light curves were excluded from the analysis. We accounted for correlated residual red noise following the methodology of \cite{pont06}. In summary, the method models the binned variance with the relation $\sigma^2 = \sigma_w^2/N + \sigma_r^2$ relation, where $\sigma_w$ is the uncorrelated white noise component, $N$ is the number of measurements in the bin and $\sigma_r$ is the red noise component. {\rm{We found white noise dispersion in the range from about 350 to 750 parts-per-million. For the red noise, we found a dispersion in the range from about 30 to 50 parts-per-million.}} Results for the WFC3 transmission spectrum are plotted in Figures\,\ref{fig:fig0A} to \ref{fig:fig2}.

\subsection{Spitzer IRAC photometry}\label{sec:irac_lcs}
Our light curve analysis of the 3.6 and $4.5\mu$m IRAC transits and eclipses is similar to the approach for the WFC\,3 spectroscopic light curves and our previous {\it{Spitzer}} studies \citep{nikolov15, sing16}. We analysed the primary/secondary eclipses and instrumental systematics simultaneously. We used the analytic formulae of \citet{mandel02} to fit the transit light curves and assumed the non-linear limb darkening law with coefficients computed from the three-dimensional stellar atmosphere model grid of \cite{magic2015}. To obtain an analytical eclipse model, we relied on the same formulae, but computed model with limb darkening set to zero. We fixed the remaining system parameters to literature values listed in Table\,\ref{tab:systpartab} and fitted only for the eclipse depth and central time. To correct for the intrapixel sensitivity-induced flux variations we fit a polynomial function of the stellar centroid position of the form: 
\begin{equation}
f(t)=a_0 +a_1x+a_2x^2 +a_3y+a_4y^2 +a_5xy+a_6t,
\label{eqn:irac_model}
\end{equation}

where $f(t)$ is the stellar flux as a function of time, $t$; $x$ and $y$ are the positions of the stellar centroid on the detector and $a_0$ to $a_6$ are the free parameters of the fit. We produced all possible combinations of terms from Equation\,\ref{eqn:irac_model} and performed separate fits. Using the AIC statistic, we marginalized the models to obtain parameters with uncertainties following \citet{gibson14}.

\subsubsection{Primary transits}
First, we fit the {\it{Spitzer}} transit light curve with the goal to measure the planet system parameters. We allowed the planet orbital inclination, $i$, normalized semi-major axis, $a/R_{\ast}$, planet-to-star radius ratio, $R_{p}/R_{\ast}$ and central transit time, $T_0$ to vary free in the fit. The planet orbital period and eccentricity were fixed to the values reported in Table\,\ref{tab:systpartab} and the four limb darkening coefficients were held fixed for the $3.6\mu$m and $4.5\mu$m as follows: $c_1=0.5493$, $c_2=-0.4770$, $c_3=0.4054$, $c_4=-0.1443$ and $c_1=0.5480$, $c_2=-0.6455$, $c_3=0.6196$, $c_4=-0.2232$ respectively. We found inclination values of $i=84.75\pm0.75^{\circ}$ and $85.35\pm0.41^{\circ}$ and normalized semi-major axis of  $a/R_{\ast} = 8.48\pm0.71$ and $a/R_{\ast} = 8.99\pm0.42$ for the $3.6\mu$m and $4.5\mu$m transits respectively. We combine these measurements with other results to refine the system parameters and planet orbital ephemeris in Section\,\ref{sec:res_syspar}.

To obtain measurements for the transmission spectrum, we first allowed the mid-transit time (${T_0}$) and planet-to-star radius ($R_p/R_\ast$) to vary and held fixed the orbital period ($P$), eccentricity($e$), inclination and scaled semi-major axis ($a/R_\ast$) to the values reported in Table\,\ref{tab:systpartab}. We found central transit times ${\rm T_{mid}}=2458779.05194\pm0.00061$ and $2458785.90231\pm0.00045$ (${\rm{BJD_{TDB}}}$) for the 3.6 and $4.5\mu$m channel, respectively. Finally, we repeated the fit allowing only the planet radius to vary and fixed the mid-transit times to their best-fit values from the first fit. Table\,\ref{tab:transpec} and Figure\,\ref{fig:figA3a} summarize our results. Applying  the wavelet method on the light curve residuals, as detailed in \citep{carter09}, we found white and red noise components $\sigma_w = 0.010$ and $\sigma_r = 0.0010$ for the $3.6\mu$m transit data, and $\sigma_w = 0.012$ and $\sigma_r = 0.0010$ for the $4.5\mu$m transit, respectively.

\subsubsection{Secondary eclipses}
First, we fit the secondary eclipses by allowing the mid-eclipse time (${T_{ecl}}$) and eclipse (or occultation) depth, $\delta_{occ}=(B_{\lambda}(T_{p})/B_{\lambda}(T_{\ast}))(R_{p}/R_{\ast})^2$ to vary freely. We find ${T_{ecl}(3.6)}=2458787.6139\pm0.0033$ and ${T_{ecl}(4.5)}=2458794.4777\pm0.0075$. An analysis of the departures of these measurements from the predicted ephemeris is presented in Section\,\ref{sec:res_syspar}. We measure white and red noise components of $\sigma_w = 0.010$ and $\sigma_r = 0.003$ for the $3.6\mu$m transit data, and $\sigma_w = 0.013$ and $\sigma_r = 0.0013$ for the $4.5\mu$m transit, respectively. 

In a second pass, we fixed the central eclipse times to these values and allowed only the occultation depths to vary in the light curve fits. We find $\delta_{occ}(3.6) = 1600\pm250$ and $\delta_{occ}(4.5) = 926\pm334$ parts-per-million (Figure\,\ref{fig:fig4}).

\subsection{TESS photometry}\label{sec:tess_lcs}
We performed fits to the TESS light curves using the same method as for the {\it{Spitzer}} light curves. We assumed the following limb darkening coefficients: $c_1 =   0.6228$, $c_2 =  -0.1870$, $c_3 =   0.4742$, $c_4 =  -0.2224$ and kept them fixed throughout the analysis. In addition, the planet orbital eccentricity and period were held fix to the values reported in Table\,\ref{tab:systpartab}. We first allowed all transit parameters to vary freely in the fit aiming to measure the system parameters. A weighted mean to the orbital inclination and normalized semi-major axis measured from the 2018 and 2020 light curves gave $i=85.11\pm0.27^{\circ}$ and  $a/R_\ast=8.66\pm0.26$. These values are in excellent agreement with the VLT results,  Table\,\ref{tab:systpartab}. Results for the central times are detailed in Table\,\ref{tab:oc}. In a second pass, we fixed the transit central times to the best-fit values and allowed only the planet-to-star radius ($R_p/R_\ast$) to vary. Combining all measured TESS radii, we find a weighted mean value of $R_p/R_\ast = 0.11986\pm0.00061$.

\begin{table}
\centering
\caption{Stellar elemental abundances of WASP-96}
\label{tab:stelab}
\begin{tabular}{lr} % four columns, alignment for each
\hline
\hline
\vspace{-0.05cm}
Abundance & Value \\
\hline
${\rm{[C/H]}}$ & $+0.25 \pm 0.19$ \\
${\rm{[N/H]}}$ & $+0.32 \pm 0.16$ \\
${\rm{[O/H]}}$ & $+0.37 \pm  0.11$ \\
${\rm{[Na/H]}}$ & $+0.31 \pm 0.13$ \\
${\rm{[K/H]}}$ & $+0.27 \pm 0.19$ \\
${\rm{log A(Li)}}$ & $1.69 \pm 0.15$ \\
${\rm{[Mg/H]}}$ & $+0.47 \pm 0.10$ \\
${\rm{[Ca/H]}}$ & $+0.18 \pm 0.16$ \\
${\rm{[Sc/H]}}$ & $+0.14 \pm 0.21$ \\
${\rm{[Ti /H]}}$ & $+0.18 \pm 0.12$ \\
${\rm{[V/H]}}$ & $+0.22 \pm 0.15$ \\
${\rm{[Cr/H]}}$  & $+0.25 \pm 0.11$ \\
${\rm{[Mn/H]}}$ & $+0.49 \pm 0.28$ \\
${\rm{[Fe/H]}}$ & $+0.21 \pm 0.15$ \\
${\rm{[Co/H]}}$ & $+0.20 \pm 0.15$ \\
${\rm{[Ni/H]}}$ & $+0.24 \pm 0.15$ \\
${\rm{[Y/H]}}$ & $+0.17 \pm 0.23$ \\
${\rm{[Si/H]}}$ & $+0.24 \pm 0.16$ \\
${\rm{[S/H]}}$ & $+0.25 \pm 0.11$ \\
\hline
\end{tabular}
\end{table}

\subsection{Stellar elemental abundances, chromospheric activity, photometric variability}\label{sec:tess_lcs}
We obtained abundances of the main atmospheric constituents of the host star using the FEROS spectra, using methods similar to those given in \cite{doyle13}. Results from our analysis are shown in Table\,\ref{tab:stelab}. Abundances are given with respect to the Solar abundances from \cite{asplund09}, except for lithium which is given as logarithm of the number ratio with respect to hydrogen plus 12 (e.g. \citealt{2014dapb.book..245B}, p.247).

We compared the tabulated and observed wavelengths of the {\mbox{Ca\,\sc{II}\,H\&K}} line cores finding evidence for an offset of the K line wavelength as indicated in Figure\,\ref{fig:figHK}. This has been observed toward other exoplanet host stars observed with FEROS, including WASP-110 and can be attributed to the relatively low signal-to-noise ratio \citep{2021AJ....162...88N}. We measure the activity index, $S_{\rm HK}$, as estimated from the spectrum using the method described by \cite{1978PASP...90..267V} and transformed to the $\log R'_{\rm HK}$ system using the relations from \cite{1984ApJ...279..763N}. We find the star to be  chromospherically quiet with ($\log{\rm{R^{'}_{H\&K}}}=-5.3\pm0.1$). This places WASP-96 among some of the most chromospherically quiet stars hosting a transiting exoplanet.

To assess the level of photometric variability due to spots on the host star, we performed Lomb-Scargle periodogram analysis on the ASAS-SN light curve with trial frequencies ranging between 0.005 and $1\,\rm{d^{-1}}$, corresponding to a period range between 1 and 200\,d. We find a rotational modulation with a period of $17.8$\,d with a false alarm probability (FAP) of $\sim0.1$. We phase-folded the light curve and fitted four-parameter sine wave to measure the amplitude of the flux variation. We found an amplitude of $0.49\pm0.11\%$ and residual scatter of 11 parts-per-thousand (ppt), Figure\,\ref{fig:asassn_folded}. This amplitude exceeds the $<1$mmag ($95\%$), or $<0.1\%$, rotational modulation reported by \cite{hellier14} by a factor of $\sim5$. However, given the low value for chromospheric activity index in combination with the low photometric variability from the WASP photometry we conclude that the ASAS-SN result should be taken with a low weight. The TESS SAP PDC corrected light curves from sectors 2 and 29 have an average scatter of 3.4 ppt. They were examined using {\sc period04} \citep{2005CoAst.146...53L}. No variations above 0.3 ppt were found at periods longer than the orbital period of the system. Therefore, these data confirm the lack of variation found in the WASP photometer and rule out the variation suggested by the ASAS-NS data.

%{\bf{TESS should also be able to inform on this.  You can easily take the photometry, and bin the light curve to be able to see ~mmag oscillations or not.}}

%\cite{hellier14} report $2\sim5$ Gyr age for WASP-96 from Lithium and gyrochronologic age of $8^{+26}_{-8}$. 
%Our result agrees with the 4\,mmag ($0.4\%$) upper limit for flux variability of WASP-110 reported by \cite{Anderson2014}. \cite{Anderson2014} also found that WASP-110 is an evolved $8.6\pm3.5$\,Gyr G9 star, which implies a gyrochronologic rotation comparable or exceeding the equatorial rotation period of the Sun i.e., $\gtrsim25$\,days. \cite{Anderson2014} report $v\sin(i_{\ast})=0.2\pm0.6$\,${\rm{kms^{-1}}}$, which translates to a rotation period $P_{\ast}(i_{\ast}=90^{\circ})\gtrsim34$\,days. While this is a discrepancy with our finding, a detailed comparison is hampered by the unconstrained stellar axial tilt, which can significantly decrease the actual rotation period. It is worth mentioning that empirical relations of the rotation periods to the chromospheric activity level $\log{R'_{\rm HK}}$ predict $\sim22$\,days for WASP-110 \citep[their Figure 13]{2015MNRAS.452.2745S}. Given the spread of $\sim10$\,days in the empirical relationship for the rotation period for G-type stars with $-0.1<\left[{\rm{Fe/H}}\right]<0.1$, the predicted estimate is consistent with our measured period. This is also the case for $v\sin(i_{\ast}\sim4^{\circ})$ of \cite{Anderson2014}. }} 

%------------------------------------------------------------------------------------------------------
\section{Results and interpretation }\label{sec:res}
%------------------------------------------------------------------------------------------------------
\subsection{HST-Spitzer transmission spectrum}
The measured {\it{HST}} and {\it{Spitzer}} wavelength-dependent relative planet radii, which comprise the planet's transmission spectrum are shown in Figure\,\ref{fig:fig0A}. The {\it{HST}} spectrum exhibits the full absorption signature of the rotational-vibrational absorption band of water with peaks at $\sim0.95$, $1.15$ and $1.4\mu$m covering $\sim7$ atmospheric pressure scale heights (with 1 scale height corresponding to $\sim610$\,km, assuming an equilibrium temperature of $T_{{\rm{eq}}}=1285$\,K, surface gravity of $g=7.6\,\mathrm{m/s^2}$ and solar mean molecular weight of $\mu=2.3$ a.m.u.), in a wavelength region from $\sim0.8$ to $1.7\mu$m.

\subsection{Combining VLT with HST and Spitzer transmission spectroscopy}
Following the ultimate goal of the present study i.e., extending the optical ground-based transmission spectrum of WASP-96b to infrared wavelengths, we complemented the VLT spectrum with the {\it{HST-Spitzer}} transit observations analyzed here. Figure\,\ref{fig:fig0B} shows the combined ground-based and space-based spectra, presented in \cite{2018Natur.557..526N} and in this work, respectively. The infrared spectrum is found to be at a substantially larger radius of $\Delta R_{{\rm p}}/R_{\ast}\sim 0.004$, or a change of transit depth of $\sim0.1\%$ ($1000$ ppm). Factors with the potential to contribute to such spectral offsets include: 1) stellar activity and photometric variability of the host star; 2) discrepancy in the system parameters assumed in the light curve fits of the separate data sets; 3) accurate accounting of the limb-darkening; 4) unaccounted systematic errors in the light curve analysis. The $<0.1\%$ photometric variability of the host star reported by \cite{hellier14}, or even the higher variability of $0.49\pm0.11\%$ tentatively seen in the ASAS-SN monitoring, can only account for a negligible radius offset. In this study, we assumed identical system parameters for each of the data set and the limb darkening was properly treated in both studies. We note that the limb darkening of the host star is predicted to substantially vary from optical to infrared wavelengths. The VLT data have been fit with a quadratic limb darkening law  with the first coefficient allowed to vary to account for the variable transmission of the Earth atmosphere. In this study we fixed the limb darkening coefficients to their theoretical values from the same synthetic spectra assumed for the limb darkening of the ground-based data \citep{magic2015}. A visual inspection of the light curve fits shown in Extended Data Figures 2 \& 3 of \cite{2018Natur.557..526N} show no evidence for significant unaccounted systematics. However in that study (and here too), the authors performed a common mode correction, which strongly depends on the assumption for system parameters and particularly on the planet-to-star radius measurement. \cite{2018Natur.557..526N} measured planet-to-star radii of $R_{{\rm p}}/R_{\ast}=0.1147\pm0.0014$ and $0.1168^{+0.0015}_{-0.0014}$ for the GRIS600B and GRIS600RI white light curves, respectively. While these values differ by only $\Delta R_{{\rm p}}/R_{\ast} =0.0021\pm0.0021$, the choice for either of them displaces the mean radius of the transmission spectrum by a substantial margin. The authors chose to rely on the blue (GRIS600B) transmission spectrum, which is less affected by telluric lines, and made use of the overlapping region of the two grisms to offset the red (GRIS600RI) transmission spectrum to the inherent for the blue (GRIS600RI) spectrum smaller planet radii. Choosing the red spectrum as a reference would account for $50\%$ of the difference between the {\it{HST}} and VLT spectra. While \cite{2021AJ....161....4Y} concluded that combining datasets is risky and could not ascertain if the VLT and {\it{HST}} data were compatible, we point out that techniques such as common-mode corrections are well known to compromise the absolute transit depths, such that an overall transmission spectrum can have a large vertical uncertainty which is not fully captured by the point-to-point uncertainties of the relative transit depths (typically the quantity of interest). This is largely because the common-mode correction effectively assumes a broadband $R_{{\rm p}}/R_{\ast}$ with no uncertainty, which ignores the actual uncertainty on the broadband $R_{{\rm p}}/R_{\ast}$ from the white light curve fit.

%These details are discussed in \cite{2018Natur.557..526N}, yet are not mentioned in the \cite{2021AJ....161....4Y} study as a source of the offset and disagreement.

With a TESS measurement from a broad-band covering the wavelength region from 0.6 to $1\mu$m, one could employ this radius to offset the VLT spectrum. However, the TESS band is centered in the small overlapping region of the red (GRIS600RI) VLT and blue {\it{HST}} (G102) data, which hampers quantifying the size of the offset. We also constructed a band in the {\it{HST}} G102 data corresponding to the narrow overlapping region with the VLT data. However, we found that the uncertainty of such a narrow band of only 3nm is insufficient to derive an accurate radius. Instead, we derive an offset correction for the VLT data by a simultaneous retrieval of the VLT and {\it{HST-Spitzer}} spectra as detailed in the next section.

\begin{table*}
\centering
\caption{WASP-96b VLT, {\it{HST}} and {\it{Spitzer}} transmission spectrum after a correction of the VLT spectrum \label{tab:transpec}}
\label{tab:transpec}
\begin{tabular}{ccccccccc}
\hline
\hline
\vspace{-0.05cm}
$\mathrm{\lambda_c}$, $\mathrm{\mu}$m & $\mathrm{\Delta\lambda, \mu}$m & $\mathrm{R_p/R_s}$ & $\mathrm{\sigma_{R_p/R_s}}$  & & $\mathrm{\lambda_c}$, $\mathrm{\mu}$m & $\mathrm{\Delta\lambda, \mu}$m & $\mathrm{R_p/R_s}$ & $\mathrm{\sigma_{R_p/R_s}}$\\
\hline
0.37565	&	0.02565	&	0.11908	&	0.00152	&	&	1.12345	&	0.00465	&	0.11741	&	0.00120	\\
0.40530	&	0.00400	&	0.11876	&	0.00182	&	&	1.13275	&	0.00465	&	0.11813	&	0.00124	\\
0.41330	&	0.00400	&	0.11811	&	0.00167	&	&	1.14205	&	0.00465	&	0.11972	&	0.00111	\\
0.42130	&	0.00400	&	0.11983	&	0.00167	&	&	1.15135	&	0.00465	&	0.12159	&	0.00116	\\
0.42930	&	0.00400	&	0.11888	&	0.00205	&	&	1.16065	&	0.00465	&	0.12027	&	0.00114	\\
0.43730	&	0.00400	&	0.11750	&	0.00163	&	&	1.16995	&	0.00465	&	0.11739	&	0.00123	\\
0.44530	&	0.00400	&	0.11934	&	0.00129	&	&	1.17925	&	0.00465	&	0.11790	&	0.00114	\\
0.45330	&	0.00400	&	0.11895	&	0.00123	&	&	1.18855	&	0.00465	&	0.11724	&	0.00127	\\
0.46130	&	0.00400	&	0.11891	&	0.00125	&	&	1.19785	&	0.00465	&	0.11794	&	0.00100	\\
0.46930	&	0.00400	&	0.11790	&	0.00116	&	&	1.20715	&	0.00465	&	0.11980	&	0.00105	\\
0.47730	&	0.00400	&	0.11711	&	0.00100	&	&	1.21645	&	0.00465	&	0.11834	&	0.00106	\\
0.48530	&	0.00400	&	0.11753	&	0.00130	&	&	1.22575	&	0.00465	&	0.11900	&	0.00129	\\
0.49330	&	0.00400	&	0.11870	&	0.00104	&	&	1.23505	&	0.00465	&	0.11600	&	0.00122	\\
0.50130	&	0.00400	&	0.11880	&	0.00138	&	&	1.24435	&	0.00465	&	0.11965	&	0.00088	\\
0.50930	&	0.00400	&	0.11755	&	0.00112	&	&	1.25365	&	0.00465	&	0.11953	&	0.00105	\\
0.51730	&	0.00400	&	0.11747	&	0.00101	&	&	1.26295	&	0.00465	&	0.11730	&	0.00092	\\
0.52530	&	0.00400	&	0.11746	&	0.00096	&	&	1.27225	&	0.00465	&	0.11929	&	0.00095	\\
0.53330	&	0.00400	&	0.11873	&	0.00089	&	&	1.28155	&	0.00465	&	0.12127	&	0.00126	\\
0.54130	&	0.00400	&	0.11794	&	0.00095	&	&	1.29085	&	0.00465	&	0.11891	&	0.00117	\\
0.54930	&	0.00400	&	0.11790	&	0.00089	&	&	1.30015	&	0.00465	&	0.11765	&	0.00128	\\
0.55730	&	0.00400	&	0.11751	&	0.00102	&	&	1.30945	&	0.00465	&	0.11952	&	0.00129	\\
0.56530	&	0.00400	&	0.11970	&	0.00095	&	&	1.31875	&	0.00465	&	0.11893	&	0.00124	\\
0.57330	&	0.00400	&	0.11928	&	0.00095	&	&	1.32805	&	0.00465	&	0.12073	&	0.00109	\\
0.58130	&	0.00400	&	0.12082	&	0.00084	&	&	1.33735	&	0.00465	&	0.11894	&	0.00094	\\
0.58930	&	0.00400	&	0.12114	&	0.00092	&	&	1.34665	&	0.00465	&	0.11730	&	0.00136	\\
0.59730	&	0.00400	&	0.12077	&	0.00086	&	&	1.35595	&	0.00465	&	0.11908	&	0.00139	\\
0.60530	&	0.00400	&	0.12036	&	0.00086	&	&	1.36525	&	0.00465	&	0.12145	&	0.00117	\\
0.61330	&	0.00400	&	0.11975	&	0.00081	&	&	1.37455	&	0.00465	&	0.12052	&	0.00106	\\
0.62130	&	0.00400	&	0.11916	&	0.00096	&	&	1.38385	&	0.00465	&	0.12164	&	0.00136	\\
0.62930	&	0.00400	&	0.12077	&	0.00099	&	&	1.39315	&	0.00465	&	0.12144	&	0.00118	\\
0.63730	&	0.00400	&	0.11886	&	0.00096	&	&	1.40245	&	0.00465	&	0.12082	&	0.00097	\\
0.64530	&	0.00400	&	0.11928	&	0.00091	&	&	1.41175	&	0.00465	&	0.11989	&	0.00109	\\
0.65330	&	0.00400	&	0.11982	&	0.00100	&	&	1.42105	&	0.00465	&	0.12124	&	0.00125	\\
0.66130	&	0.00400	&	0.11713	&	0.00089	&	&	1.43035	&	0.00465	&	0.11941	&	0.00106	\\
0.66930	&	0.00400	&	0.11961	&	0.00081	&	&	1.43965	&	0.00465	&	0.12087	&	0.00135	\\
0.67730	&	0.00400	&	0.11871	&	0.00117	&	&	1.44895	&	0.00465	&	0.12023	&	0.00097	\\
0.68930	&	0.00800	&	0.11618	&	0.00078	&	&	1.45825	&	0.00465	&	0.11984	&	0.00122	\\
0.70130	&	0.00400	&	0.11740	&	0.00090	&	&	1.46755	&	0.00465	&	0.11852	&	0.00099	\\
0.70930	&	0.00400	&	0.11912	&	0.00096	&	&	1.47685	&	0.00465	&	0.12035	&	0.00099	\\
0.71730	&	0.00400	&	0.11857	&	0.00085	&	&	1.48615	&	0.00465	&	0.11982	&	0.00113	\\
0.72530	&	0.00400	&	0.11892	&	0.00097	&	&	1.49545	&	0.00465	&	0.12122	&	0.00094	\\
0.73330	&	0.00400	&	0.12001	&	0.00094	&	&	1.50475	&	0.00465	&	0.11976	&	0.00136	\\
0.74130	&	0.00400	&	0.11786	&	0.00081	&	&	1.51405	&	0.00465	&	0.11897	&	0.00112	\\
0.74930	&	0.00400	&	0.11834	&	0.00096	&	&	1.52335	&	0.00465	&	0.11942	&	0.00095	\\
0.76130	&	0.00800	&	0.11902	&	0.00075	&	&	1.53265	&	0.00465	&	0.11795	&	0.00128	\\
0.77330	&	0.00400	&	0.11741	&	0.00109	&	&	1.54195	&	0.00465	&	0.11757	&	0.00113	\\
0.78130	&	0.00400	&	0.11771	&	0.00096	&	&	1.55125	&	0.00465	&	0.11615	&	0.00105	\\
0.78930	&	0.00400	&	0.11765	&	0.00123	&	&	1.56055	&	0.00465	&	0.11766	&	0.00127	\\
0.79730	&	0.00400	&	0.11716	&	0.00117	&	&	1.56985	&	0.00465	&	0.11889	&	0.00112	\\
0.82460	&	0.03430	&	0.11857	&	0.00099	&	&	1.57915	&	0.00465	&	0.11759	&	0.00124	\\
0.87600	&	0.01710	&	0.11682	&	0.00079	&	&	1.58845	&	0.00465	&	0.11823	&	0.00115	\\
0.91025	&	0.01715	&	0.11868	&	0.00074	&	&	1.59775	&	0.00465	&	0.11736	&	0.00091	\\
0.94455	&	0.01715	&	0.11833	&	0.00076	&	&	1.60705	&	0.00465	&	0.11872	&	0.00139	\\
0.97885	&	0.01715	&	0.11860	&	0.00069	&	&	1.61635	&	0.00465	&	0.11842	&	0.00129	\\
1.01320	&	0.01720	&	0.11864	&	0.00072	&	&	1.62565	&	0.00465	&	0.11829	&	0.00140	\\
1.04755	&	0.01715	&	0.11822	&	0.00080	&	&	1.63495	&	0.00465	&	0.11817	&	0.00130	\\
1.08185	&	0.01715	&	0.11693	&	0.00104	&	&	3.60000	&	0.35460	&	0.12140	&	0.00130	\\
1.11415	&	0.00465	&	0.11829	&	0.00101	&	&	4.50000	&	0.45020	&	0.11890	&	0.00120	\\
1.11615	&	0.01715	&	0.11984	&	0.00063	&	&		&		&		&		\\
\hline
\end{tabular}
\end{table*}

\begin{figure}
\centering
\includegraphics[width=0.99\linewidth]{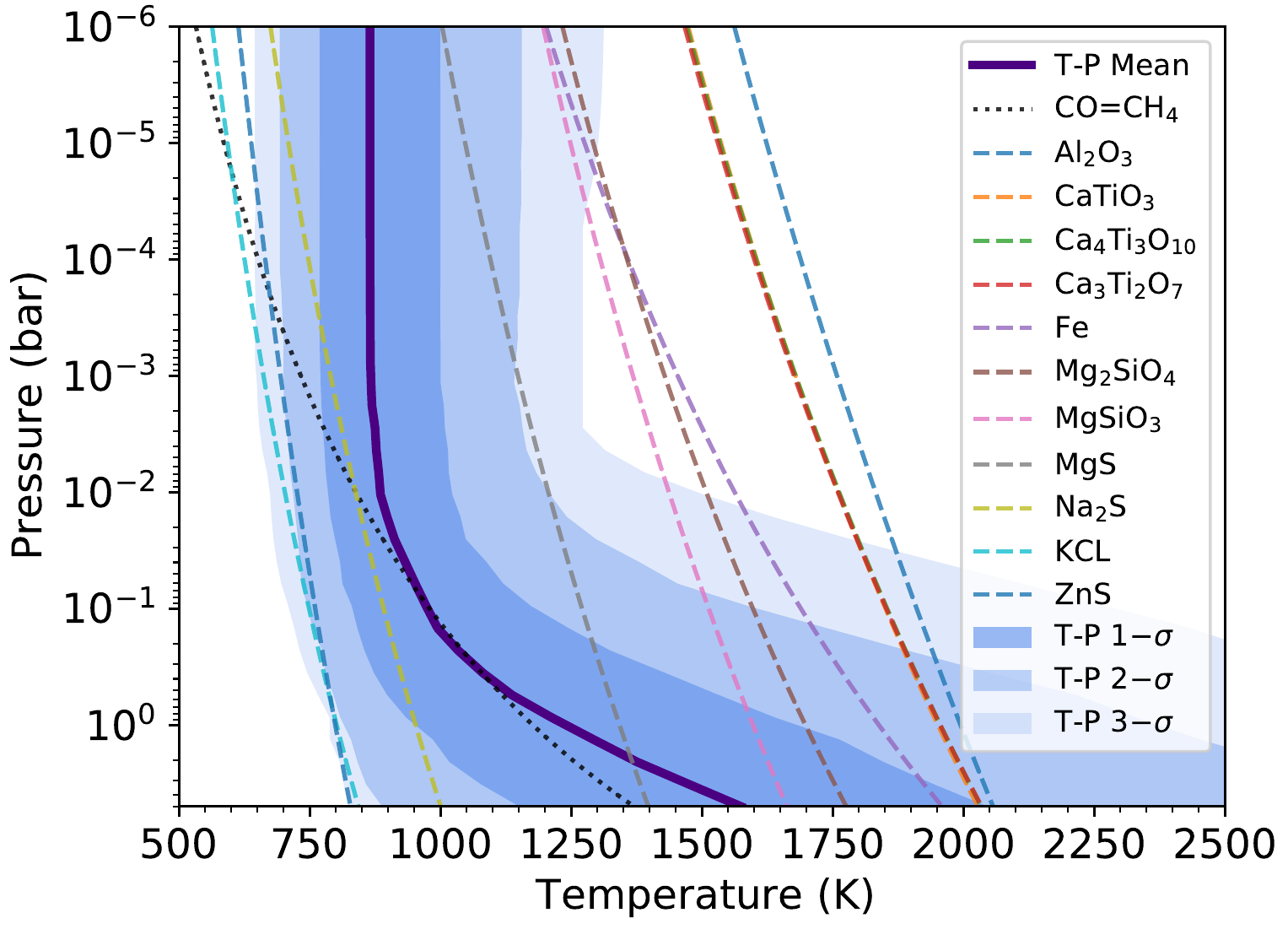}
\caption{Pressure-temperature profile results from {\tt{ATMO}} assuming chemical equilibrium. Shown are the median profile of the posterior distribution (purple line) and the $1\sigma$, $2\sigma$ and $3\sigma$ model distributions with dark, medium and light blue shaded regions, respectively. Cloud condensation curves, assuming solar abundance are indicated with the dashed lines from~\protect\citealt[][]{visscher06, 2010ApJ...716.1060V, 2017MNRAS.464.4247W}.}
\label{fig:pt}
\end{figure}

\subsection{Retrieval of the combined VLT, HST and Spitzer transmission spectrum}
We considered inverse methods to constrain atmospheric properties at the day-night terminator of WASP-96b and the radius difference between the VLT optical and {\it{HST-Spitzer}} infrared transmission spectra. We employed the 1D-2D radiative-convective equilibrium {\tt{ATMO}} model {\citep{tremblin15, tremblin16}}. Previous retrieval results using {\tt{ATMO}} can be found in \cite{2017Sci...356..628W, evans17, 2019MNRAS.488.2222M, 2020MNRAS.496.1638M, 2021MNRAS.500.4042S, 2018AJ....156..283E}. {\tt{ATMO}} is capable of solving for the pressure-temperature (P-T) profile and chemical abundances that satisfy hydrostatic equilibrium and conservation of energy given a set of opacities. The code computes isotropic multi-gas Rayleigh scattering, H$_2-$H$_2$ and H$_2-$He collision-induced absorption, as well as opacities for all major chemical species taken from the most up-to-date high-temperature line list sources, including Na, K, Li, Rb, Cs, H$_2$, He, H$_2$O, CO$_2$, CO, CH$_4$ {\citep{1984ApJS...56..193S, 2008JPhCS.130a2011H, 2006MNRAS.368.1087B, 2011JQSRT.112.1403T, 2010JQSRT.111.2139R, 2014MNRAS.440.1649Y, tremblin15, tremblin16, amundsen14, amundsen17, drummond16, goyal17}}. Rainout of condensate species is also included (see discussion in \citealt{2019MNRAS.486..783G}). {\tt{ATMO}} can also fit a parameterized P-T profile and chemical abundances to transmission spectra. For the retrieval analysis in our study, we made use of the parameterized P-T profile detailed in \cite{guillot10}, which gives three free parameters: the Planck mean thermal infrared opacity ($\kappa_{\mathrm{IR}}$); the ratio of optical to infrared opacities ($\gamma_{\mathrm{O/IR}}$); and an irradiation efficiency factor ($\beta$). We set the internal temperature to $400$\,K based on \cite{2019ApJ...884L...6T}.

{\tt{ATMO}} treats hazes as parameterized enhanced Rayleigh scattering and clouds are assumed to add grey opacity. The code makes use of the correlated-k approximation with the random overlap method to compute the total gaseous mixture opacity, which has been shown to agree well with a full line-by-line treatment \citep{amundsen17}.

We first performed a retrieval assuming chemical equilibrium and allowed the radius, opacity from clouds and hazes and elemental abundances of Na, K, C and O allowed to freely vary. In the synthetic spectra, alkali line-wing shapes were computed assuming the predictions from \cite{allard12}. We allowed four elemental abundances to vary independently, as they are major species that are also likely to be sensitive to spectral features in the data, while the rest were varied by a trace metallicity parameter $\log{(Z_{\mathrm{trace}}/Z_{\odot})}$. This quantity is not the overall bulk metallicity but contains the abundances for trace species not otherwise individually fit (i.e., all except H, He, Na, K, C and O). By varying the elemental abundances of Na, K, C and O separately, we allow for non-solar compositions but with chemical equilibrium imposed such that each model fit has a chemically plausible mix of molecules given the retrieved temperatures, pressures, and underlying elemental abundances. Importantly, by varying both O and C separately (rather than a single C/O value) we alleviate an important modeling assumption that can affect the retrieved C/O value \citep[see][]{2019MNRAS.486.1123D}. To quantify the radius difference between the VLT and {\it{HST-Spitzer}} spectra, we included a parameter that controls the offset of the VLT spectrum ($\delta_{\mathrm{VLT}}$) and was allowed to freely vary in the retrieval. We chose to discard the TESS radius measurement owing to its large uncertainty and broad wavelength region. 

\begin{figure*}
\centering
\includegraphics[trim={1.2cm 0.0cm 0.7cm 0.0cm}, width=0.98\linewidth]{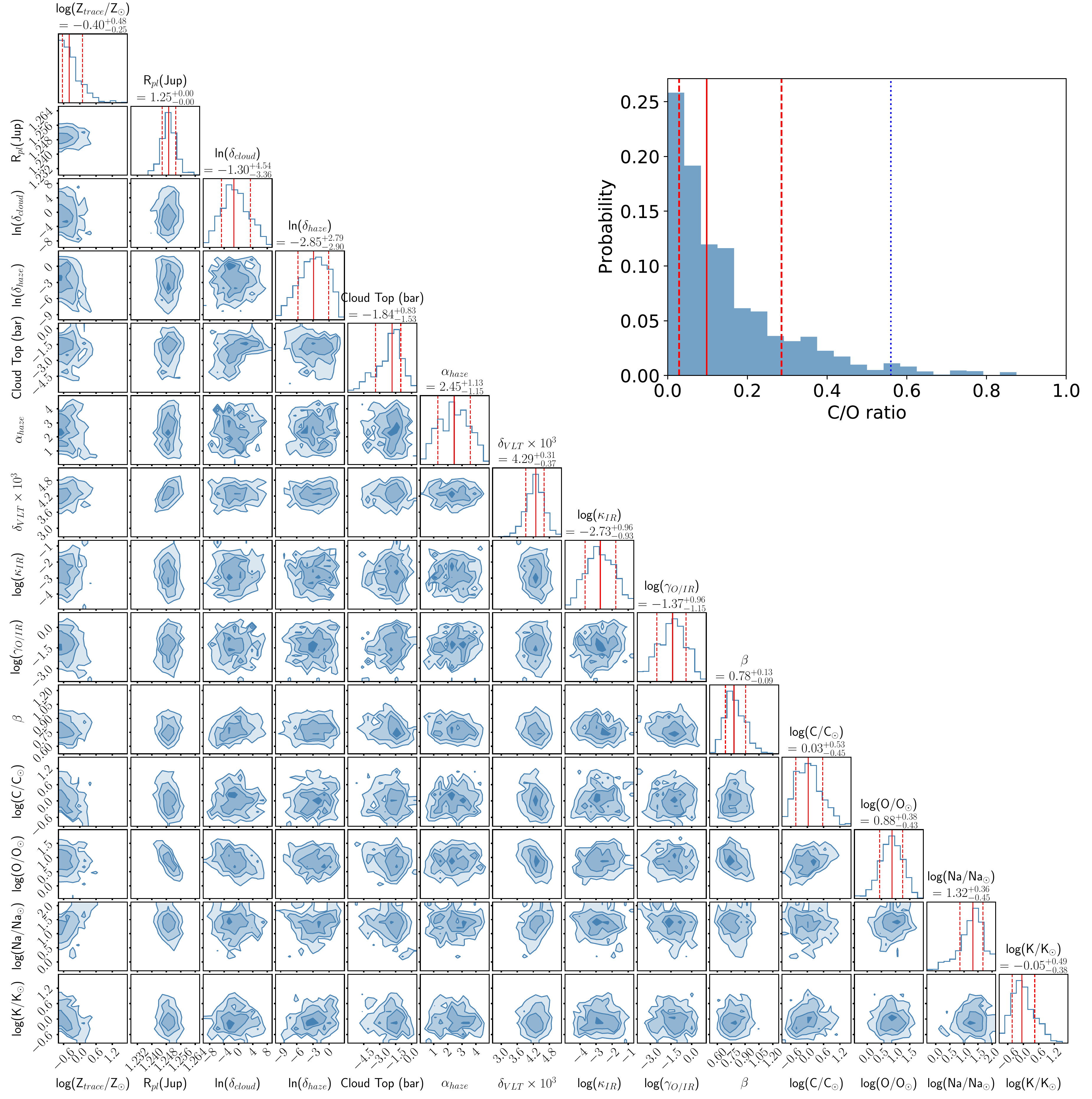}
\caption{Atmospheric properties for WASP-96b obtained from a retrieval with {\tt{ATMO}}, assuming chemical equilibrium. Marginalized posterior distributions of trace metallicity, planet radius, cloud and haze opacity, cloud top pressure, index of haze power, infrared opacity, optical-to-infrared offset, $\beta$ and elemental abundances of carbon, oxygen, sodium and potassium in solar units. Lack of clouds and hazes at the pressures being probed by our observations is evidenced by the high value of cloud top pressure and negligible haze opacity of the marginalized posterior distributions. The abundances of oxygen and sodium are the two constrained quantities. The mean abundances and $1\sigma$ uncertainties are indicated with the vertical continuous and dashed lines, respectively. The posterior distribution of the C/O ratio is reconstructed from the $\mathrm{\log(C/C_{\odot})}$ and $\mathrm{\log(O/O_{\odot})}$ abundances and is shown, with the solar value indicated with the blue dotted line in the top panel.}
\label{fig:figretr}
\end{figure*}

The best-fit transmission spectra are shown in Figure\,\ref{fig:fig1} and \ref{fig:fig2}. The P-T profile is shown in Figure\,\ref{fig:pt}; and the posteriors of all retrieved quantities are shown in Figure\,\ref{fig:figretr}. Our retrieval analysis finds a negligible contribution from cloud or haze opacity, which indicates that the atmosphere of WASP-96b is cloud and haze-free at the pressures being probed at the limb. This result is in agreement with the retrieval results from our VLT optical spectrum alone detailed in \cite{2018Natur.557..526N}. We find an offset in the planet-to-star radius of $\delta_{\mathrm{VLT}}=0.0043^{+0.00031}_{-0.00037}$ (or a transit depth change of 1004 ppm), which is in agreement with the results of \cite{2021AJ....161....4Y}.

We obtain constraints of $\mathrm{log(Na/Na_{\odot})}=1.32^{+0.36}_{-0.45}$ for the sodium abundance, $\mathrm{log(K/K_{\odot})}=-0.05^{+0.49}_{-0.39}$ for the potassium abundance, $\mathrm{log(C/C_{\odot})}=0.03^{+0.53}_{-0.45}$ for the carbon abundance, and $\mathrm{log(O/O_{\odot})}=0.88^{+0.38}_{-0.43}$ for the oxygen abundance, Table\,\ref{tab:retres} and Figure\,\ref{fig:figretr}. Sodium and Oxygen are the sole two elements for which we place definitive constraints, which correspond to solar to supersolar values with abundances relative to the solar values of {\mbox{$21^{+27}_{-14}$ (2.4-$\sigma$) and $7^{+11}_{-4}$ (2-$\sigma$)}}, respectively. Our results are in agreement with the measured supersolar abundances of the host star Table\,\ref{tab:stelab}. The best-fit model gives $\chi^2=120.96$ for 103 degrees of freedom. We note that the posterior distribution of the C/O ratio shown in Figure\,\ref{fig:figretr} has been reconstructed from the best-fit retrieved $\mathrm{log(C/C_{\odot})}$ and $\mathrm{log(O/O_{\odot})}$ abundances. This implies that the prior ranges for the C/O ratio are constructed for the $\mathrm{log(O/O_{\odot})}$ and $\mathrm{log(C/C_{\odot})}$ parameters which are fit separately. One limitation of the data and retrieval modeling is that the carbon-bearing species aren’t resolved by the {\it{Spitzer}} data, including CO$_2$, CO and CH$_4$. While our modeling assumes equilibrium chemistry to model the mixture of these gasses, it has previously been found the C/O ratio can vary dramatically depending on whether free chemistry or equilibrium chemistry assumed in the model (e.g., \citealt{2021MNRAS.500.4042S}).

\begin{table}
\centering
\caption{Retrieval results}
\label{tab:retres}
\begin{tabular}{lcc} % four columns, alignment for each
\hline
\hline
\vspace{-0.05cm}
Parameter, Unit & Value & Prior range \\
\hline
$\chi^2_{\mathrm{min}}$ & $120.96$ & \\
$N_{\mathrm{free}}$ & 14 & \\
$N_{\mathrm{data}}$ & 117 & \\
$\log(Z_{\mathrm{trace}}/Z_{\odot})$ & $-0.40^{+0.48}_{-0.25}$ & $-1$ to $2$\\
$R_{\mathrm{pl}}$, ($R_{\mathrm{Jup}}$) & $1.2493^{+ 0.0039}_{-0.0036}$ & $1.13$ to $1.39$\\
$\ln{\delta_\mathrm{cloud}}$ & $-1.30^{+4.54}_{-3.36}$ & $-10$ to $10$\\
$\ln{\delta_\mathrm{haze}}$ & $-2.85^{+2.79}_{-2.90}$ & $-10$ to $10$\\
Cloud Top, $\log_{10}{\mathrm{(bar)}}$ & $-1.84^{+0.83}_{-1.53}$ & $-6$ to $0.7$\\
$\alpha_{\mathrm{haze}}$ & $2.45^{+1.13}_{-1.15}$ & $0$ to $5$\\
$\delta_{\mathrm{VLT}}\times10^{3}$ & $4.29^{+0.31}_{-0.37}$ & $0.29$ to $8.3$\\
$\log{\kappa_\mathrm{IR}}$ & $-2.73^{+0.96}_{-0.93}$ & $-5$ to $0.5$\\
$\log{\gamma_\mathrm{O/IR}}$ & $-1.37^{+0.96}_{-1.15}$ & $-4$ to $1.5$\\
$\beta$ & $0.78^{+0.13}_{-0.09}$ & $0$ to $2$\\
$\mathrm{\log(C/C_{\odot})}$ & $0.03^{+0.53}_{-0.45}$ & $-1$ to $2$\\
$\mathrm{\log(O/O_{\odot})}$ & $0.88^{+0.38}_{-0.43}$ &  $-1$ to $2$\\
$\mathrm{\log(Na/Na_{\odot})}$ & $1.32^{+0.36}_{-0.45}$ &  $-1$ to $2$\\
$\mathrm{\log(K/K_{\odot})}$ & $-0.05^{+0.49}_{-0.39}$ &  $-1$ to $2$\\
\hline
\end{tabular}
\end{table}

% $\mathrm{[K/K_{\odot}]}=-0.05^{+0.50}_{-0.39}$
% $\mathrm{[O/O_{\odot}]}=0.87^{+0.38}_{-0.43}$
% $\mathrm{[Na/Na_{\odot}]}=1.32^{+0.36}_{-0.46}$
% $\mathrm{[C/C_{\odot}]}=0.03^{+0.53}_{-0.45}$

%log(C/C$\tex_{sun}$)     0.032807249 1-sig:     -0.45086149      0.53228100  2-sig:     -0.77353400       1.0754330  3-sig:     -0.95859551       1.4342847
%log(O/O$\tex_{sun}$)      0.87426026 1-sig:     -0.43156137      0.37944120  2-sig:     -0.86961185      0.69552372  3-sig:      -1.2665272      0.97566825
%log(Na/Na$\tex_{sun}$)       1.3201057 1-sig:     -0.46126640      0.35746868  2-sig:      -1.1665475      0.69546015  3-sig:      -1.5331998      0.79715619
%log(K/K$\tex_{sun}$)    -0.050290142 1-sig:     -0.38507785      0.49508374  2-sig:     -0.74454305       1.0302661  3-sig:     -0.83486048       1.4831406

In addition to the retrieval with alkali metals profile from \cite{allard12}, we performed a second retrieval assuming Lorentzian profile from \cite{2000ApJ...531..438B}. Our results indicate an excellent agreement between the retrieved atmospheric properties of the two assumed profiles without a preference of one versus the other.

We also performed retrievals excluding the {\it{Spitzer}} data.  Because little-to-no carbon species are needed to fit the transmission spectrum (with low $\mathrm{log(C/C_{\odot})}$  and resulting C/O ratio found), and the {\it{Spitzer}} data points have fairly large error bars, excluding the {\it{Spitzer}} data had little effect on the retrieved parameters.  The $\mathrm{log(C/C_{\odot})}$  parameter was found to increase by about 0.1 dex when excluding the {\it{Spitzer}} data, but this is well within the retrieval uncertainties (0.5 dex). 

The combined VLT-{\it{HST}}-{\it{Spitzer}} transmission spectrum after a correction of the VLT spectrum is detailed in Table\,\ref{tab:transpec}.

%\cite{visscher06, 2010ApJ...716.1060V} and and \cite{2017MNRAS.464.4247W}

\subsection{Forward Model analysis of the combined VLT, HST and Spitzer transmission spectrum}
To interpret the combined {\mbox{WASP-96b}} observed transmission spectrum we also used a forward model grid of transmission spectra generated using self-consistent P-T profiles in {\tt{ATMO}} for a range of metallicity, C/O ratio and energy re-distribution, described in detail in \citet{2020MNRAS.498.4680G}. Here, self-consistent imply radiative-convective equilibrium P-T profiles  consistent with equilibrium chemical abundances. Prior to fitting the data, we offset the VLT spectrum with by $\delta=4.29$, found in the retrieval analysis. The best fit forward model transmission spectrum is shown in Figure \ref{fig:forwd}. This fit corresponds to $\chi^2 = 129.3$ and reduced $\chi^2 = 1.11$ with 116 degrees of freedom (117 observed points minus 1 for the vertical offset in the transmission spectrum). This fit could be considered as a relatively good fit given that the model transmission spectrum is purely forward with self-consistent P-T profiles. The best fit transmission spectrum has a re-circulation factor of 0.25 (high energy redistribution), indicating a relatively cooler P-T profile as expected for transmission spectrum, since we probe the day-night terminator region of the exoplanet using the transmission spectrum. We also determine that the best fit forward model transmission spectrum has solar metallicity and sub-solar C/O ratio (0.35). Only one other model spectrum from the grid lie within the $1\sigma$ $\chi^2$ value of the best-fit transmission spectrum and this model also has solar metallicity and sub-solar C/O ratio (0.35).

\subsection{Thermal emission spectrum}
The two measured {\it{Spitzer}} IRAC eclipse depths are plotted in Figure\,\ref{fig:fig4} and comprise the planet's dayside thermal spectrum. We find the 4.5$\mu$m eclipse depth to be lower compared to the $3.6\mu$m data, which suggests a decreasing temperature with altitude.

We first fit a blackbody synthetic spectrum to determine the dayside temperature of the planet. The blackbody temperature was the sole parameter allowed to freely vary in this fit. We assumed a blackbody spectrum for the host star with the effective temperature of WASP-96. We found that the best-fitting spectrum for the planet, with $\chi^2=6.7$ for 1 degree of freedom, and ${\rm{BIC=7.4}}$, corresponds to a temperature of $T_{\rm{p}}=1545\pm90$\,K. This spectrum corresponds to the dashed line in Figure\,\ref{fig:fig4}.% While higher, this measurement is only $260$K hotter compared with the planet's equilibrium temperature.

We then employed the publicly available radiative transfer code {\it{petitRADTRANS}} \citep{2019A&A...627A..67M} to compute synthetic spectra for the planet's dayside. {\it{petitRADTRANS}} computes planet emission spectra for an assumed pressure-temperature (P-T) profile and gravity. Because exoplanet emission spectra are known to depend on the underlying P-T profile, we chose to compute spectra, assuming P-T profiles of a decreasing (non-inverted), and an increasing (inverted) temperature with altitude. {\mbox{WASP-96b}} is irradiated rather modestly, and we wouldn't expect to encounter an inverted temperature profile. We note that the spectrum for the inverted case is included merely for reference purposes, not expecting it to fit the data. We made use of the analytic P-T profile of \cite{guillot10}. We assumed solar abundances for the main atmospheric constituents, including H$_2$O, CO, CO$_2$ and CO$_4$ and a solar mean molecular weight of 2.33. Both spectra were computed assuming unity for the recirculation factor i.e., no redistribution of the input stellar energy in the planet's atmosphere. Prior to comparing synthetic with observed data, we averaged the synthetic spectra within the wavelength bins of the observed spectrum and computed the corresponding $\chi^2$. We find the observed 3.6 and $4.5\mu$m eclipse depths to be best explained by a spectrum assuming the non-inverted profile with a chi-square of $\chi^2=4.6$. The spectrum assuming an inverted profile resulted in  $\chi^2=8.8$. We note that complementary observations at a higher signal-to-noise ratio and spectral resolution would be needed to confidently confirm this result.

We also interpreted the {\it{Spitzer}} IRAC eclipse depths with a self-consistent planet-specific forward model grid of emission spectra as shown in \citet{2021ApJ...923..242G}. These emission spectra were generated using radiative-convective equilibrium P-T profiles consistent with equilibrium chemical abundances in {\tt{ATMO}} for WASP-96b, for a range of metallicity, C/O ratio and energy re-distribution, described in detail in \citet{2020MNRAS.498.4680G}. The best fit model emission spectrum and the corresponding P-T profile is shown in Figure \ref{fig:fig4}. We find the observed 3.6 and $4.5\,\mu$m eclipse depths to be best explained by a spectrum with non-inverted P-T profile with  $\chi^2=3.8$. The best fit model emission spectrum corresponds to a re-circulation factor of 1, implying none of the energy is redistributed to the night side and therefore the P-T profile corresponding to this best fit model emission spectrum is one of the hottest in the grid of models for WASP-96b. The best fit model emission spectrum corresponds to a metallicity of 200 times solar and C/O ratio of 0.7. However, by looking at all the model spectra that lie within the $1\sigma$ $\chi^2$ value of the best-fit emission spectrum we determine that a large range of metallicities (From 1x to 200x solar) and C/O ratio (0.35 to 0.7) can explain observed eclipse depths due to large uncertainty in the $4.5\,\mu$m eclipse depth. Interestingly, all these model spectra similar to the best-fit model spectrum have non-inverted P-T profiles and correspond to a re-circulation factor of 1 (no-redistribution), driven by the higher $3.6\,\mu$m eclipse depths requiring higher temperatures. The strong absorption in the $4.5\,\mu$m band is a result of CO, leading to lower eclipse depth in this band as seen in left side of Figure \ref{fig:fig4} for the best fit model. Therefore, the best fit model has a non-inverted P-T profile which results in this absorption feature (instead of an inverted P-T profile that would lead to an emission feature due to CO, in the $4.5\,\mu$m band). All the model spectra that lie within the $1\sigma$ $\chi^2$ value of the best-fit emission spectrum show this CO absorption feature but with varying degree of strength (i.e model $4.5\,\mu$m eclipse depth). 

\subsection{Refined system parameters}\label{sec:res_syspar}
To improve the planet orbital inclination and semi-major axis of WASP-96b, we combined our measurements with literature results. In particular, our {\it{Spitzer}} and TESS measurements were complemented with the VLT results reported in \cite{2018Natur.557..526N}. Weighted mean results for the orbital inclination, $i$ and normalized semi-major axis $a/R$ are reported in Table\,\ref{tab:spar_eph}. Both results agree within $2\sigma$ with the measurements of \cite{hellier14}.
%$i=85.135\pm0.081^{\circ}$ and  $a/R_\ast=8.820\pm0.090$. 

\subsection{Improved planet orbital ephemeris}\label{sec:res_syspar}
We combined the central transit times from our {\it{HST}}, {\it{Spitzer}} and TESS light curve analysis with the transit times reported by \cite{hellier14} and \cite{2018Natur.557..526N} to derive an updated transit ephemeris of WASP-96b. These two studies respectively include a single transit mid-time from a combined analysis of WASP, EulerCam and TRAPPIST light curves, and central times from two VLT FORS2 white light curves. All central times were converted to BJDs (by making use of the utilities detailed in \citealt{eastman10}) and fitted with a linear function of the planet orbital period ($P$), zero-epoch central transit time ($T_{0}$) and transit epoch ($E$): $T_{\rm{C}}(E)=T_{0}+EP$. Our results for the fitted parameters are reported in Table\,\ref{tab:spar_eph} and resulted in a reduced chi-square of $\chi^2_r=1.03$ for $18$ degrees of freedom. The updated ephemeris is in an excellent agreement with the ephemeris reported by \cite{hellier14} and \cite{2021AJ....161....4Y} and does not indicate the presence of transit timing variations (TTVs). Results for the measured mid-transit times, observed minus computed transit times (O$-$C) are tabulated in Table\,\ref{tab:oc} and plotted in Figure\,\ref{fig:figOC}.
%We find an orbital period of $P = 3.42525650\pm0.00000037$\,(days) and a mid-transit time of $T_{\rm{C}}(E) = 2457963.84067\pm0.00012$\,(days), with a reduced chi-square of $\chi^2_r=1.03$. 

\begin{table}
\centering
\caption{Central transit times}
\label{tab:oc}
\begin{tabular}{rlrcl} % four columns, alignment for each
\hline
\hline
\vspace{-0.05cm}
Epoch & Mid-transit time, ${\rm{BJD}}_{{\rm{TDB}}}$ & O$-$C, day & Data & Note \\
\hline
$ -498$ & $2456258.06290\pm{0.00020}$ &  $-0.000027$ & WASP & $       1$\\
$    0$ & $2457963.84120^{+0.00034}_{-0.00033}$ &  $ 0.000534$ & VLT & $       2$\\
$    7$ & $2457987.81699\pm{0.00029}$ &  $-0.000472$ & VLT & $       2$\\
$  114$ & $2458354.32081\pm{0.00094}$ &  $ 0.000902$ & TESS & $       3$\\
$  115$ & $2458357.74406\pm{0.00086}$ &  $-0.001104$ & TESS & $       3$\\
$  116$ & $2458361.17133\pm{0.00085}$ &  $ 0.000909$ & TESS & $       3$\\
$  117$ & $2458364.59511\pm{0.00076}$ &  $-0.000567$ & TESS & $       3$\\
$  119$ & $2458371.44639\pm{0.00081}$ &  $ 0.000200$ & TESS & $       3$\\
$  120$ & $2458374.87130\pm{0.00082}$ &  $-0.000147$ & TESS & $       3$\\
$  121$ & $2458378.29731\pm{0.00102}$ &  $ 0.000607$ & TESS & $       3$\\
$  148$ & $2458470.77900^{+0.00370}_{-0.00150}$ &  $ 0.000371$ & {\it{HST}} & $       3$\\
$  151$ & $2458481.05503^{+0.00041}_{-0.00046}$ &  $ 0.000632$ & {\it{HST}} & $       3$\\
$  238$ & $2458779.05194\pm{0.00061}$ &  $ 0.000226$ & {\it{Spitzer}} & $       3$\\
$  240$ & $2458785.90231\pm{0.00045}$ &  $ 0.000083$ & {\it{Spitzer}} & $       3$\\
$  329$ & $2459090.74842\pm{0.00084}$ &  $-0.001636$ & TESS & $       3$\\
$  330$ & $2459094.17473\pm{0.00093}$ &  $-0.000582$ & TESS & $       3$\\
$  331$ & $2459097.60105\pm{0.00085}$ &  $ 0.000481$ & TESS & $       3$\\
$  333$ & $2459104.45013\pm{0.00098}$ &  $-0.000952$ & TESS & $       3$\\
$  334$ & $2459107.87539\pm{0.00098}$ &  $-0.000948$ & TESS & $       3$\\
$  335$ & $2459111.30260\pm{0.00098}$ &  $ 0.001005$ & TESS & $       3$\\
\hline
\end{tabular}
\vspace{-0.8cm}
\tablecomments{1: \cite{hellier14}, 2: \cite{2018Natur.557..526N}, 3: This work}
\end{table}

We used the refined orbital ephemeris to predict the central times of the observed secondary eclipses. Significant departures of mid-eclipse times, measured as the difference ($\Delta t_{\mathrm{O-C}}$) between the observed and computed times would indicate a non-circular planet orbit. Table\,\ref{tab:spar_eph} summarizes the values for $\Delta t_{\mathrm{O-C}}(3.6)$ and $\Delta  t_{\mathrm{O-C}}(4.5)$, which agree with the predictions at $1\sigma$ and $\sim1.6\sigma$ and indicate no departures from a circular orbit. Finally, we used Equation (6) from \cite{1950ArA.....1...59W} to calculate  $e\cos{\omega}$ for the two eclipses. Table\,\ref{tab:spar_eph} details the weighted mean value from the two observations.
% We find $\Delta t(3.6)=-1.4\pm4.8$ and $\Delta t(4.5)=18\pm11$ minutes for the two eclipses

\begin{table}
\centering
\caption{System parameters and orbital ephemeris}
\label{tab:spar_eph}
\begin{tabular}{lr} % four columns, alignment for each
\hline
\hline
\vspace{-0.05cm}
Parameter, symbol, unit & Value \\
\hline
inclination, $i$,$(^{\circ})$ & $85.135\pm0.081$ \\
$a/R_{\ast}$& $8.820\pm0.090$ \\
Orbital Period, $P$, (day) & $3.42525650\pm0.00000037$ \\
Transit central time, $T_0$, (BJD) & $2457963.84067\pm0.00012$ \\
$\Delta t_{\mathrm{O-C}}(3.6)$, (min)& $-1.4\pm4.8$\\
$\Delta t_{\mathrm{O-C}}(4.5)$, (min)& $18\pm11$\\
$e\cos(\omega)$ & $-0.00790\pm0.00010$ \\
\hline
\end{tabular}
\end{table}

%{\bf{Can quote a value of $e\cos(w)$ from the lack of timing offset. Can use Lopez-Morales (2010) equation 1. A small table of where you put this, ephemeris, and such would be good.}}

%The top two best-match models assume an isothermal and thermally inverted pressure–temperature profiles. Models assuming decreasing temperature with an increasing altitude are excluded at high confidence. (b) Zoom around the the WFC3 spectrum with models. (c) Pressure–temperature profiles assumed in the calculation of the emission models. The shaded region indicates the pressures probed by our WFC3 observations and the dashed lines indicate condensation curves for atmospheric constituents CaSiO3 and TiO.

%\begin{figure*}
%\begin{subfigure}%{2\textwidth}
%  \centering
%  % include first image
%  \includegraphics[width=1\linewidth, trim = 0cm 0cm 0cm 0cm]{arc_monitor_nested_spectra_posterior.pdf}  
%  %\caption{Put your sub-caption here}
%  %\label{fig:sub-first}
%\end{subfigure}
%\begin{subfigure}%{2\textwidth}
%  \centering
%  % include second image
%  \includegraphics[width=1\linewidth]{arc_monitor_nested_spectra_posterior_lin.pdf}  
%  %\caption{Put your sub-caption here}
%  %\label{fig:sub-second}
%\end{subfigure}
%\caption{Put your caption here}
%\label{fig:fig}
%\end{figure*}

\section{Discussion}\label{sec:disc}
The sodium abundance obtained from our retrieval analysis of the combined optical-infrared transmission spectrum, ${\rm{\log(Na/Na_{\odot})}}$ = $1.32^{+0.36}_{-0.45}$, is about one order of magnitude larger, but is still compatible with the previous determination using only the VLT spectrum, $\log{\epsilon_{\mathrm{\,Na}}}=6.9^{+0.6}_{-0.4}$ i.e., ${\rm{\log(Na/Na_{\odot})}}$ = $0.66^{+0.6}_{-0.4}$ at the $\sim1.2\sigma$ level.

\subsection{A comparison with an earlier result}\label{sec:compyip}
\begin{figure}
\includegraphics[width=0.99\linewidth]{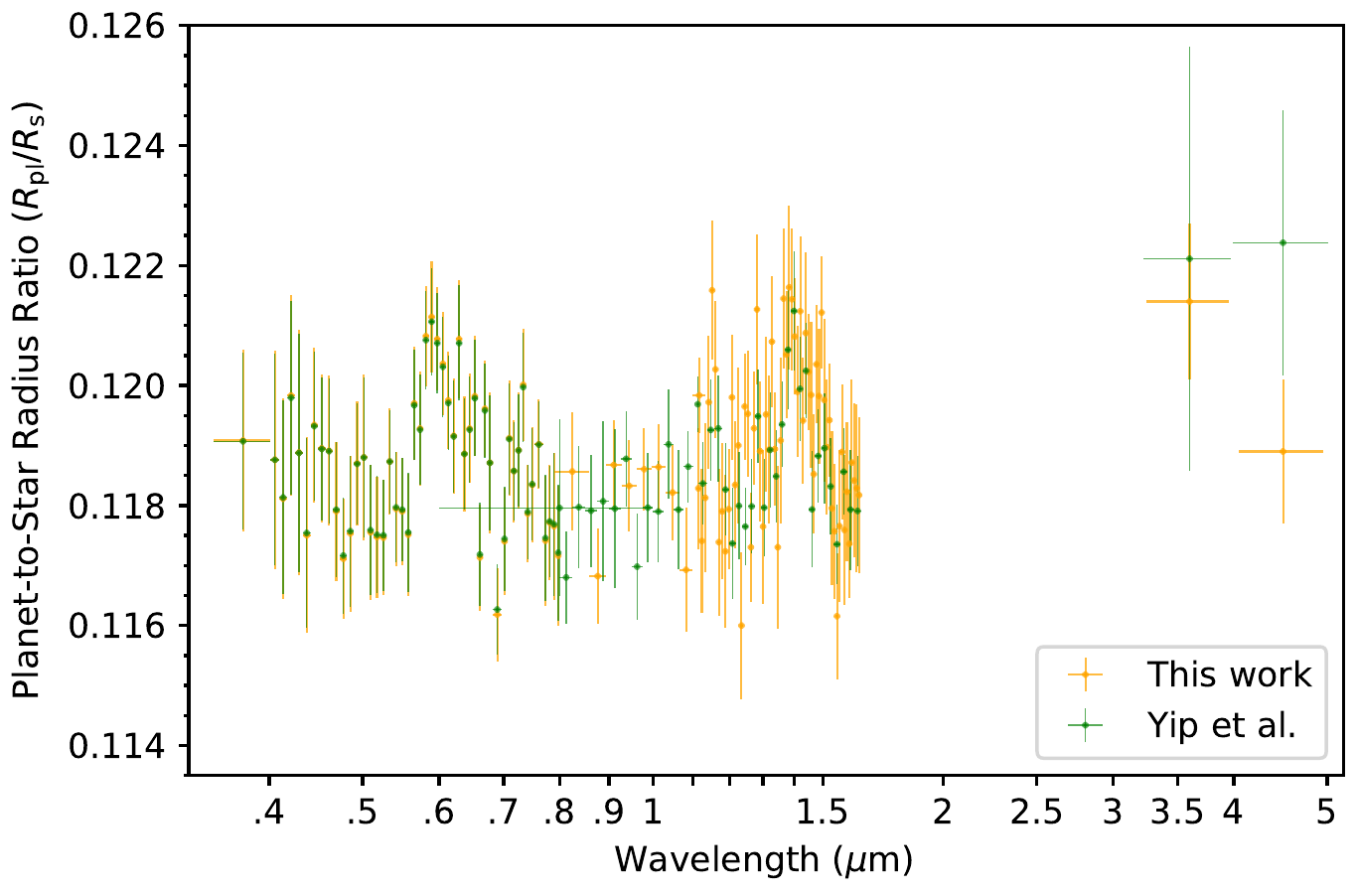}
\caption{Comparison between the result in this work with those of Yip et al. (2021). An offset of $\mathrm{\Delta R_p/R_s}=(47.36\pm0.24)\times10^{-5}$ has been added to the literature result measured using the VLT spectrum. The horizontal bars represent $1\sigma$ uncertainties and the horizontal bars are the wavelength bins used to measure the transmission spectra.  }
\label{fig:yip_comp}
\end{figure}

An analysis of the {\it{HST}} G102 and G141 transit spectroscopy and the Spitzer IRAC 3.6 and $4.5\mu$m photometric transits has been presented by \cite{2021AJ....161....4Y}. We find the two spectra to be slightly offset by a mean difference between the spectrum from this work and the literature spectrum of $\mathrm{\Delta R_p/R_s}=(47.36\pm0.24)\times10^{-5}$, measured using the VLT wavelengths. This offset is consistent the uncertainty of $\mathrm{R_p/R_s}$ found in our analysis of the white light curve (see Table\,\ref{tab:systpartab}). After applying the mean difference, the two spectra are in a respectable agreement with regards to the spectral shape (Figure\,\ref{fig:yip_comp}). The {\it{Spitzer}} IRAC radii are found to be in a good agreement at the $\sim1\sigma$ confidence level. We note significantly smaller uncertainties, a factor of $\sim2.2$ of the {\it{Spitzer}} radii measured in our work compared to the results of \cite{2021AJ....161....4Y}. Factors that can contribute to such significant difference include unaccounted systematics effects or fits performed to binned light curves. Yet, we are unsure of the exact reasons. We additionally note a marginal difference in the radii of the VLT spectrum and their uncertainties reported by \cite{2021AJ....161....4Y}. Finally, the G102 bin widths of the literature result reported in their Table\,6 are a factor of ten smaller than the plotted bin widths throughout that study. We are unsure for the reason that led to this misquotation of the published results. Despite the differences between the literature transmission spectrum and the spectrum reported in this work, both studies find abundance results in excellent agreement using different analysis and retrieval methods.

%{\bf{Compare abundance results and system parameters and orbital ephemeris}}

% You could certainly cite the volatiles to refractories ratio paper of Lothringer et al. (2021), since we have O (which is mostly thought of a volatile) and Na, which is a refractory, although the most volatile of the refractories!
%2021ApJ...914...12L

\subsection{Exoplanet abundances in the Solar System context}\label{sec:mmd}
\begin{figure}
\includegraphics[width=0.99\linewidth]{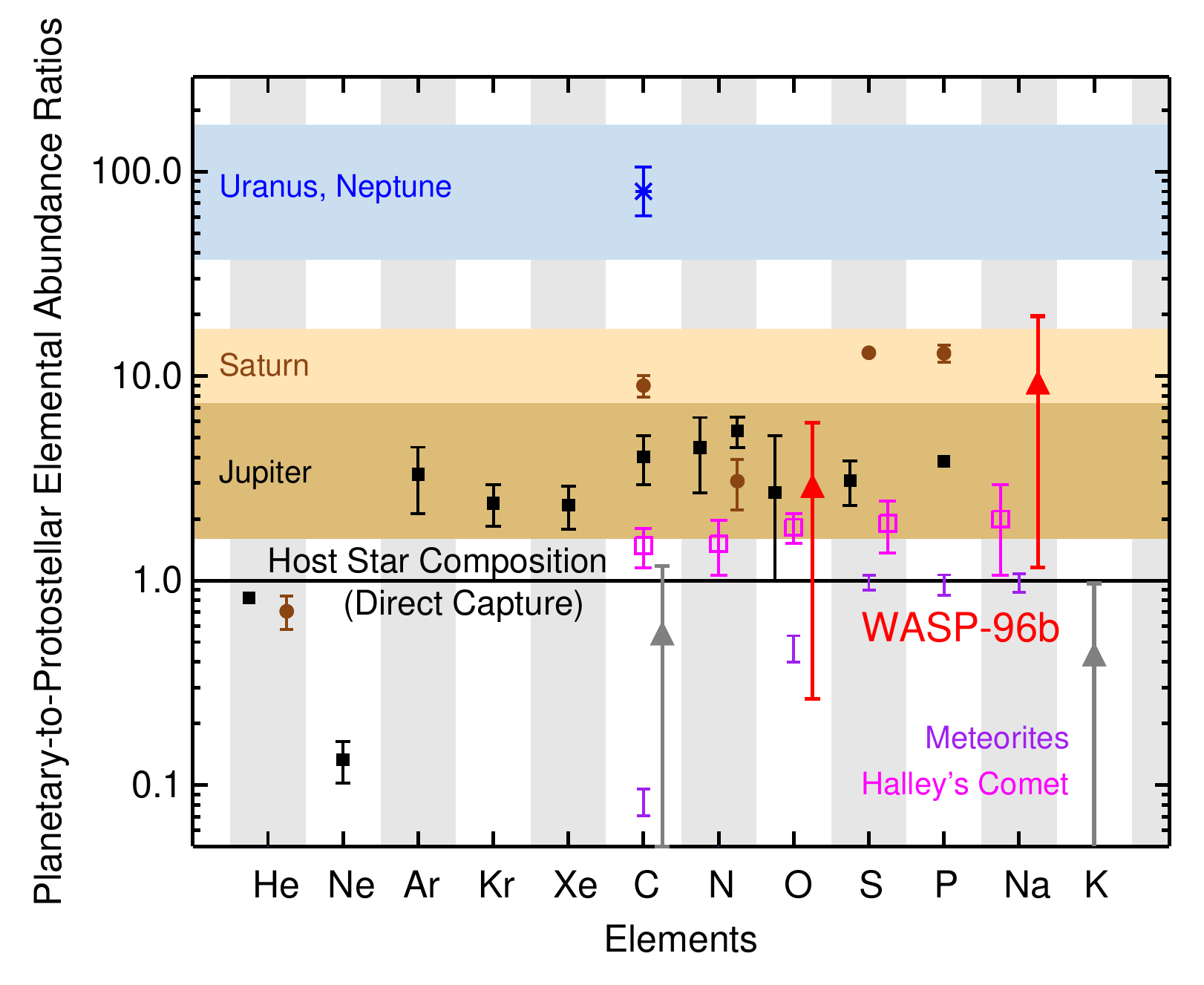}
\caption{Elemental abundance ratios in the atmospheres of our own Solar System planets Jupiter, Saturn, Uranus and Neptune, meteorites and the Halley's comet, relative to the protosolar values (adapted from~\protect\citealt[][]{2020SSRv..216...18A}). The oxygen abundance of Jupiter has been updated to the latest measurement reported by ~\protect\cite{2020NatAs...4..609L}. Elemental abundance ratios for the atmosphere of WASP-96b relative to the protstellar abundances have been obtained with the same assumption as for the protosun i.e., by increasing the protostellar value by $+0.04$ ~\protect\citep{asplund09}. Oxygen and sodium are the only two abundances obtained for WASP-96b (red symbols) and the grey triangles indicate unbounded abundances for potassium and carbon. }
\label{fig:atreya}
\end{figure}

Elemental abundances of planetary atmospheres are a consequence of the joint influence of atmospheric chemical processes, formation conditions and migration \citep{2021Natur.598..580L, 2013ApJ...775...80F, 2012ApJ...758...36M, 2021ApJ...914...12L}. In the standard core-accretion paradigm, as they form, giant planets accrete H/He-dominated gas, as well as planetesimals that enrich their H/He envelopes in metals. Higher metal enrichment is expected for low-mass H/He envelopes, as these envelopes have smaller amount of gas for the in-falling metals to be mixed into. This is the case of the gas giants Jupiter and Saturn and the ice giants Uranus and Neptune in our own Solar System, which have elemental abundances that increase with a decreasing mass and also coincides with an increasing distance from the Sun (Figure\,\ref{fig:atreya}, \citealt{2016arXiv160604510A, 2020SSRv..216...18A, 1989GeCoA..53..197A, asplund09}).

Our abundance and retrieval analysis show that both the host star and planet in the WASP-96 system have solar-to-supersolar elemental abundances of their atmospheres (Table\,\ref{tab:stelab} and \ref{tab:retres}). While metal enrichenment of planets does not appear to correlate with the metallicity of the parent star \citep{2019AJ....158..239T}, the case of WASP-96 is a prime example for a significant enrichment of both, which likely reflects the initial formation conditions of this system. Comparing the oxygen abundance ratio for the atmosphere of WASP-96b relative to its host star, we find an oxygen ratio of $3.2\pm0.4$, which is similar to the oxygen ratio of Jupiter's atmosphere and the Halley's comet in our Solar System relative to the oxygen abundance of the Sun \citep{1989GeCoA..53..197A}. We find a higher ratio of $10^{+0.4}_{-0.5}$ for the abundance of sodium in the atmosphere of WASP-96b relative to its host star, which is in agreement with the corresponding value for the Halley's comet relative to the Sun, which doesn't hold for Jupiter. A recent study by \cite{2021DPS....5321201B} shows evidence for alkali elements in the deep atmosphere of Jupiter from the Juno microwave data pointing toward severely subsolar values. This implies a much lower abundance ratio of sodium for the atmosphere of Jupiter compared to the solar value. %Our oxygen abundance ratio for the atmosphere of WASP-96b and its star is in contrast with the low oxygen and carbon values measured for the dayside atmosphere of WASP-77Ab from high-resolution spectroscopy reported by \cite{2021Natur.598..580L}. 

%\subsection{Forward models}\label{sec:compfwd}
%\subsection{Comparison to other cloud-free exoplanets}\label{sec:comclf}
 %{\bf{TBD, Cloud-free indices? If the dayside is known to be 1500K, that means the MgSiO$_3$ clouds are likely there, as that's the temperature where they condense.  So the terminator may indeed be free of these silicate clouds.  While the temperatures could be low enough for hydrocarbon hazes, perhaps the low C/O ratio (constrained by Spitzer) means less soot precursors.  This is different than say WASP-127, which is very similar in temperature but has 1) overall higher metallicity and 2) a lot more C. Perhaps a comparison to Wasp-127 would be insightful.}} 

\section{Summary and conclusions}\label{sec:concl}
We presented {\it{HST}} WFC3 transit and {\it{Spitzer}} IRAC transit and eclipse observations of the atmosphere of the hot Saturn WASP-96b. This atmosphere has been identified by our team as cloud-free from a reconnaissance optical transmission spectroscopy made with the VLT. The ground-based spectrum exhibits the full pressure-broadened sodium profile, absorption from potassium and lithium and scattering from H$_2$ in the near-UV. Our team has motivated the {\it{HST}} WFC3 observations with the goal to extend the planet's transmission spectrum at infrared wavelengths to search for the theoretically predicted absorption from H$_2$O, CO and CO$_2$, and to constrain atmospheric composition. The secondary eclipse observations were aimed at constraining the planet's dayside temperature. The {\it{HST-Spitzer}} observations reveal the full absorption signature of water, which is an independent new evidence that the planet's atmosphere is free of clouds at the terminator and at pressures being probed. Similar to earlier analysis of the {\it{HST-Spitzer}} transits, we find the {\it{HST-Spitzer}} spectrum to be at a significantly larger planet-to-star radius relative to the VLT level. We confirm and correct for a spectral offset of $\Delta R_{{\rm p}}/R_{\ast}=(-4.29^{+0.31}_{-0.37})\,\times10^{-3}$ of the VLT data relative to the {\it{HST-Spitzer}} spectrum. This offset can be explained by the assumed radius and corresponding uncertainty for the common-mode correction of the VLT spectra, which is a well-known feature of ground-based transmission spectroscopy. We find evidence for a lack of chromospheric and photometric activity of the host star and therefore have an insufficient contribution to the offset. We measure abundances for Na and O that are consistent with solar-to-supersolar, with abundances relative to solar values of {\mbox{$21^{+27}_{-14}$ (2.4-$\sigma$) and $7^{+11}_{-4}$ (2-$\sigma$)}}, respectively. While combining transmission spectra from space and the ground isn't trivial, finding the first evidence for a cloudless atmosphere from the ground and the following independent confirmation from space-based observations at new, highly-complementary wavelengths, our results demonstrate the key role of ground-based low-resolution transmission spectroscopy, particularly the VLT with FORS2, in exploring the atmospheres of transiting exoplanets. A fit of a spectrum of a blackbody isothermal atmosphere to the thermal spectrum constrains a dayside brightness temperature of $T_{\rm{p}}$=$1545$$\pm$$90$K. While metal enrichenment of exoplanets has been shown not to correlate with the metallicity of their parent stars, the case of WASP-96 is a prime example for a significant enrichment of both the star and planet, which likely reflects the initial formation conditions of this system.

\section*{Acknowledgements}

This work is based on observations with the NASA/ESA {\it{Hubble Space Telescope}} (GO-15469, \citealt{2018hst..prop15469N}), obtained at the Space Telescope Science Institute (STScI) operated by AURA, Inc. This work is based on observations made with the Spitzer Space Telescope (\citealt{2019sptz.prop14255N}), which is operated by the Jet Propulsion Laboratory, California Institute of Technology under a contract with NASA. This work is based on observations collected at the European Organization for Astronomical Research in the Southern Hemisphere under European Southern Observatory programme 199.C-0467(H). This paper includes data collected by the TESS mission, which are publicly available from the Mikulski Archive for Space Telescopes (MAST). Funding for the TESS mission is provided by NASA's Science Mission directorate. The authors are grateful to the anonymous Referee for their thoughtful comments and suggestions, which helped to improve the manuscript. NN acknowledges support for this work by NASA through grants under the {\mbox{HST-GO-15469}} program from the STScI. The research leading to these results received funding from the European Research Council under the European Union’s Seventh Framework Programme (FP7/2007-2013)/ERC grant agreement number 336792. J.M.G. and N.J.M. acknowledge support from a Leverhulme Trust Research Project Grant. This work made use of the Python Gaussian process library George. NM acknowledges funding from the UKRI Future Leaders Scheme (MR/T040866/1), Science and Technology Facilities Council Consolidated Grant (ST/R000395/1) and Leverhulme Trust research project grant (RPG-2020-82). ALC is supported by a grant from STScI (JWST-ERS-01386) under NASA contract NAS5-03127.

%The Acknowledgements section is not numbered. Here you can thank helpful
%colleagues, acknowledge funding agencies, telescopes and facilities used etc.

%%%%%%%%%%%%%%%%%%%%%%%%%%%%%%%%%%%%%%%%%%%%%%%%%%
\section*{Data Availability}

%The inclusion of a Data Availability Statement is a requirement for articles published in MNRAS. Data Availability Statements provide a standardised format for readers to understand the availability of data underlying the research results described in the article. The statement may refer to original data generated in the course of the study or to third-party data analysed in the article. The statement should describe and provide means of access, where possible, by linking to the data or providing the required accession numbers for the relevant databases or DOIs.

Raw and calibrated {\it{Hubble Space Telescope}} spectral transit time series and {\it{Spitzer Space Telescope}} transit and eclipse time series photometry are publicly available at the Mikulski Archive for Space Telescopes (MAST; \url{https://archive.stsci.edu}) and the NASA/IPAC Infrared Science Archive (IRSA; \url{https://sha.ipac.caltech.edu/applications/Spitzer/SHA/}), respectively. TESS light curves are publicly available at the MAST archive. Calibrated and extracted high-resolution FEROS spectra are publicly available via the European Southern Observatory's Spectral Data Products Query Form (\url{http://archive.eso.org/wdb/wdb/adp/phase3_spectral/form}). Broad-band light curves are publicly available at the webpage of the All-Sky Automated Survey for Supernovae (ASAS-SN; \url{https://asas-sn.osu.edu}).



\bibliographystyle{mnras}
\bibliography{researchv2.bib} % if your bibtex file is called example.bib
%sample63.bib

%\begin{thebibliography}{99}
%\end{thebibliography}

%%%%%%%%%%%%%%%%%%%%%%%%%%%%%%%%%%%%%%%%%%%%%%%%%%

%%%%%%%%%%%%%%%%% APPENDICES %%%%%%%%%%%%%%%%%%%%%

\appendix
\section{Light curves and auxiliary parameters}
Plots of the separate time series observations along with the complementary detrending data.

\begin{figure*}
	\includegraphics[width=0.70\linewidth]{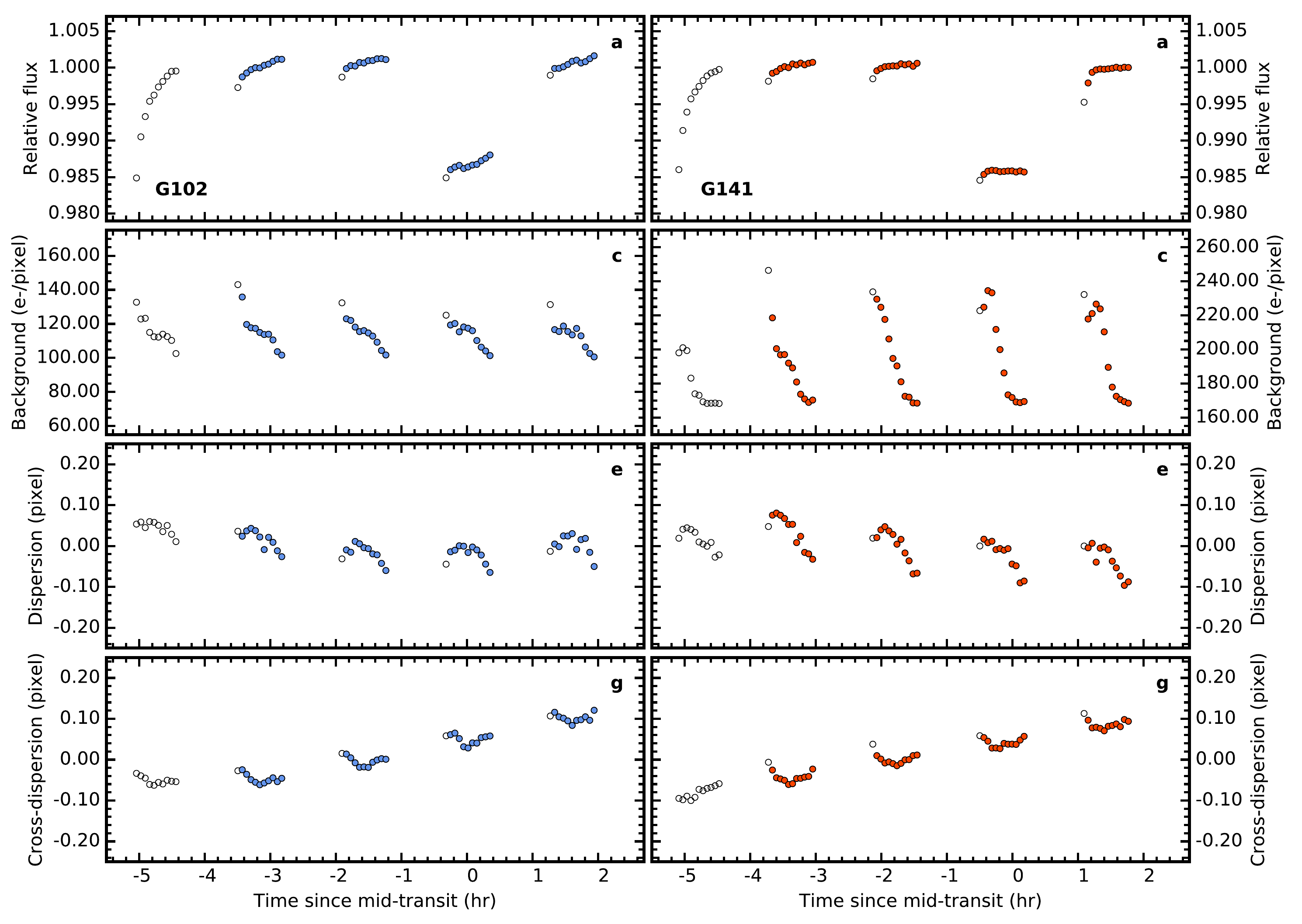}
    \caption{{\it{HST}} WFC3 white light curves and spectroscopic properties. {\it{Left}} and {\it{right}} panels refer to the G102 (blue) and G141 (red) data, respectively. Filled and open symbols, respectively, indicate retained and discarded data in our light-curve analysis. a-b, Normalised raw flux. c-d, Measured background for each frame in units of electrons per pixel, using the last individual read of each full scan. e-f, Median-subtracted drift of the spectra along the dispersion axis in units of pixels, estimated by cross-correlating the spectra with respect to the first spectrum. g-h, Same as the previous panel, but for the cross-dispersion axis measured with flux-weighted mean.}
    \label{fig:figA1}
\end{figure*}

\begin{figure*}
	\includegraphics[width=0.7\linewidth]{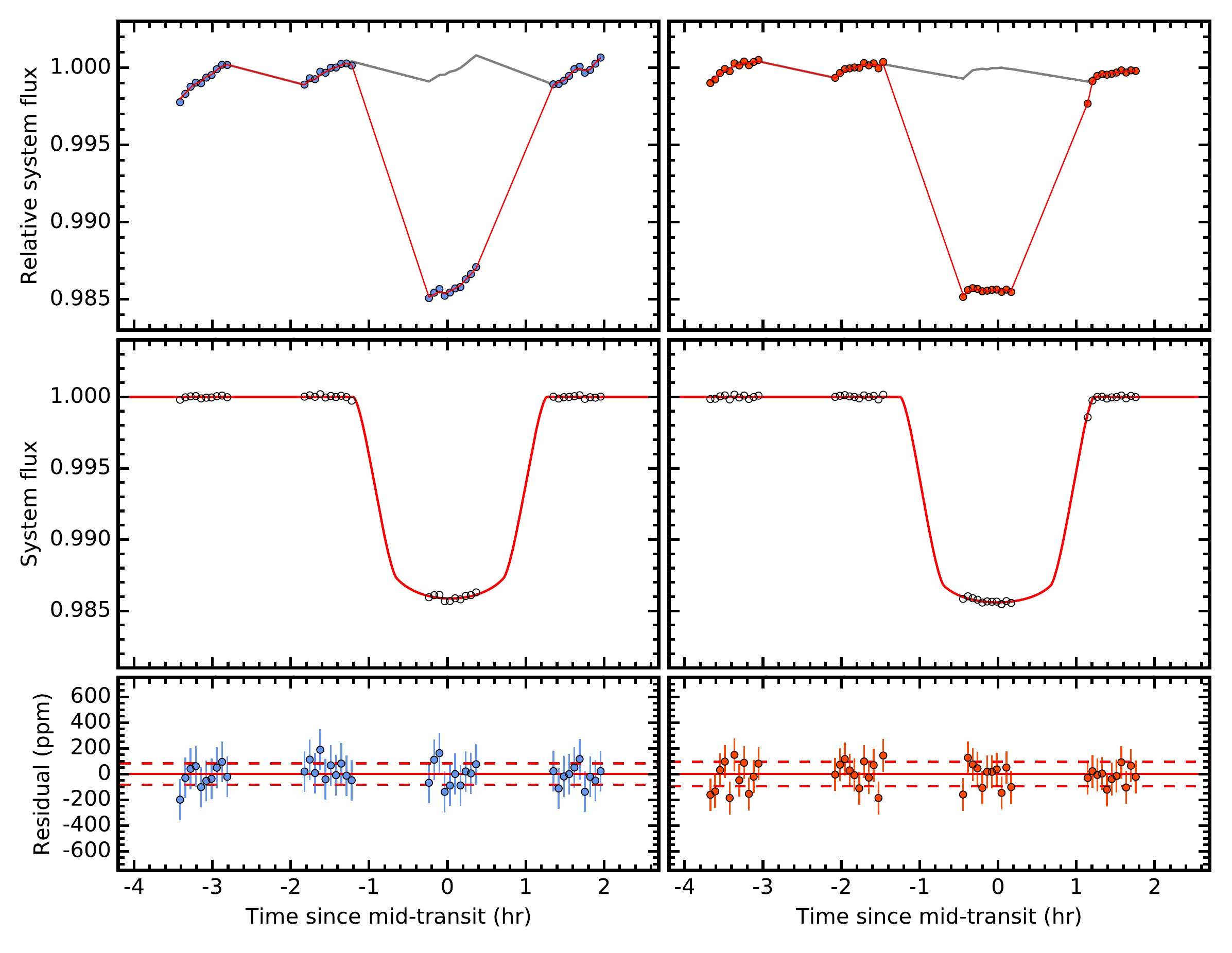}
    \caption{{\it{HST}}\,WFC3 white light curves. {\it{Left}} and {\it{right}} panels show the G102 (blue) and G141 (red) observations, respectively. The top row shows normalized raw light curves along with the marginalized Gaussian process model, assuming Matern $3/2$ kernel (grey). The later has been used as a common mode (wavelength invariant) correction for the spectroscopic light curves. The second row shows the detrended light curves along with the best-fit transit model (red). The third row shows the best-fit light curve residuals and $1\sigma$ error bars, obtained by taking the difference of the raw flux and marginalized transit and systematics models. The dashed lines indicate the light curve dispersion, $\sim75$\,parts per million.}
    \label{fig:figA2}
\end{figure*}

\begin{figure*}
	\includegraphics[width=0.85\linewidth]{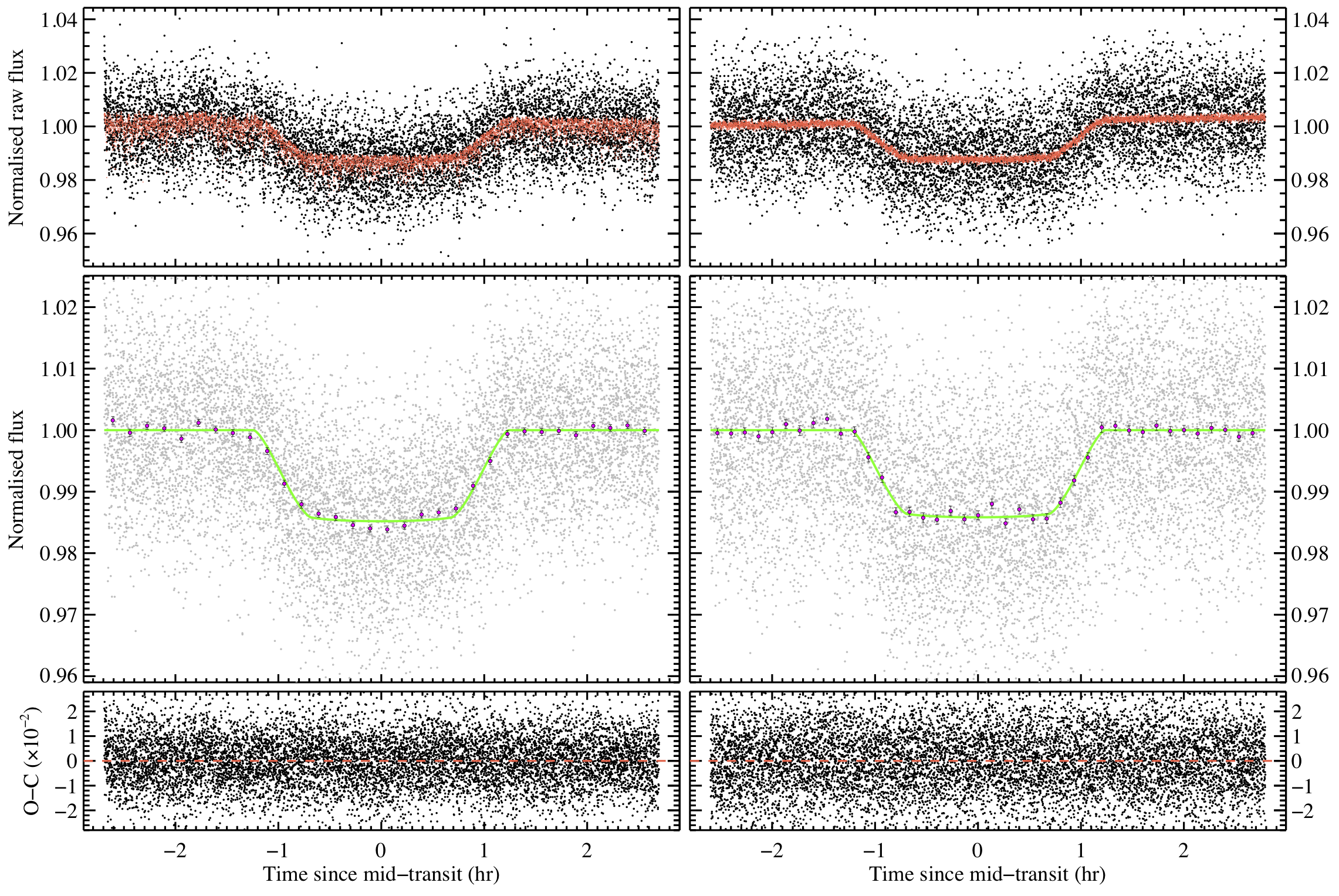}
    \caption{{\it{Spitzer}}\, IRAC 3.6 and 4.5 $\mu$m photometry (left and right,respectively).Top panels: raw flux and the best-fitting transit and systematics model; middle panels: detrended light curves and the best-fitting transit models and binned by 8 min; lower panels: light-curve residuals.}
    \label{fig:figA3a}
\end{figure*}

\begin{figure*}
	\includegraphics[width=0.85\linewidth]{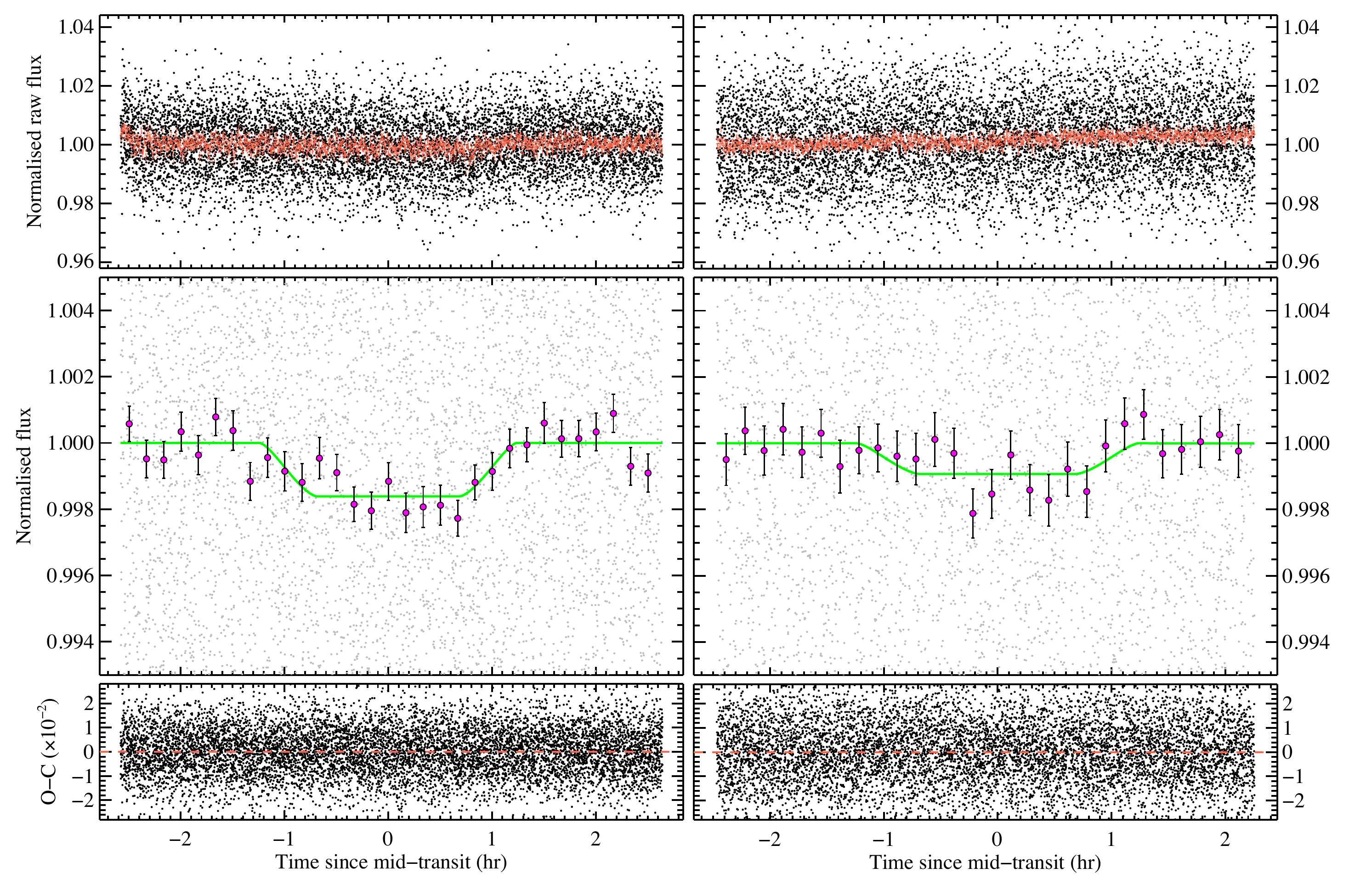}
    \caption{{\it{Spitzer}}\, IRAC 3.6 and 4.5 $\mu$m eclipse photometry (left and right,respectively).Top panels: raw flux and the best-fitting transit and systematics model; middle panels: detrended light curves and the best-fitting eclipse models and binned by 10 min; lower panels: light-curve residuals.}
    \label{fig:figA3b}
\end{figure*}

\begin{figure*}
	\includegraphics[width=0.9\linewidth]{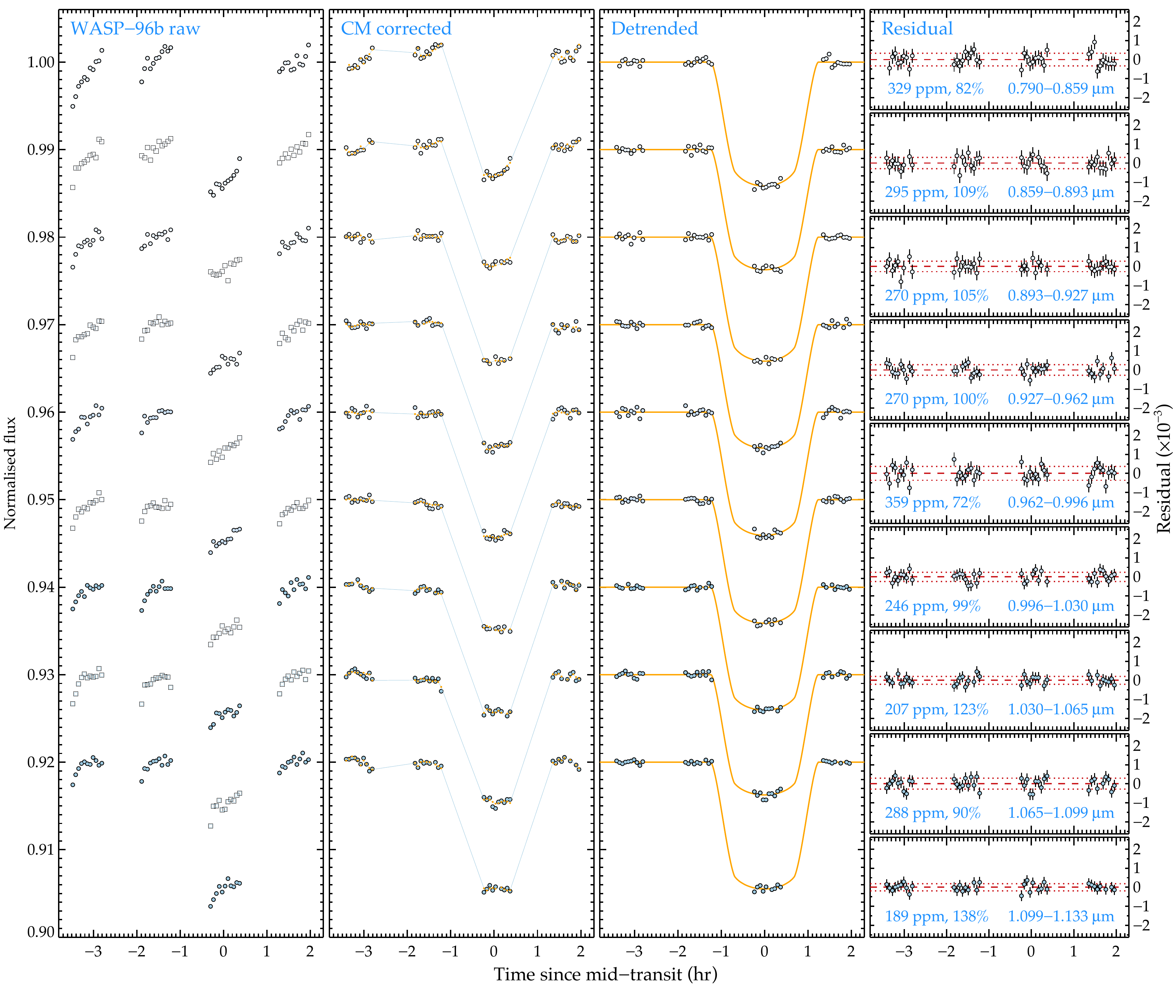}
    \caption{Spectrophotometric light curves from grism G102 offset by a constant for clarity. The first panel displays raw flux. We used dots and square symbols for the odd and even light curves, respectively to better distinguish the individual light curves. The second panel shows the common-mode (CM)-corrected light curves with transit and systematics models (lines) with the highest statistical weight. The third panel is similar to the second one, but showing the detrended light curves and transit models. The fourth panel shows residuals with $1\sigma$ uncertainties. The dashed lines indicate the median residual values, with dotted lines indicating the residual dispersion level and the percentage of the theoretical photon noise limit reached (blue) for each channel.}
    \label{fig:figA4}
\end{figure*}

\begin{figure*}
	\includegraphics[width=0.9\linewidth]{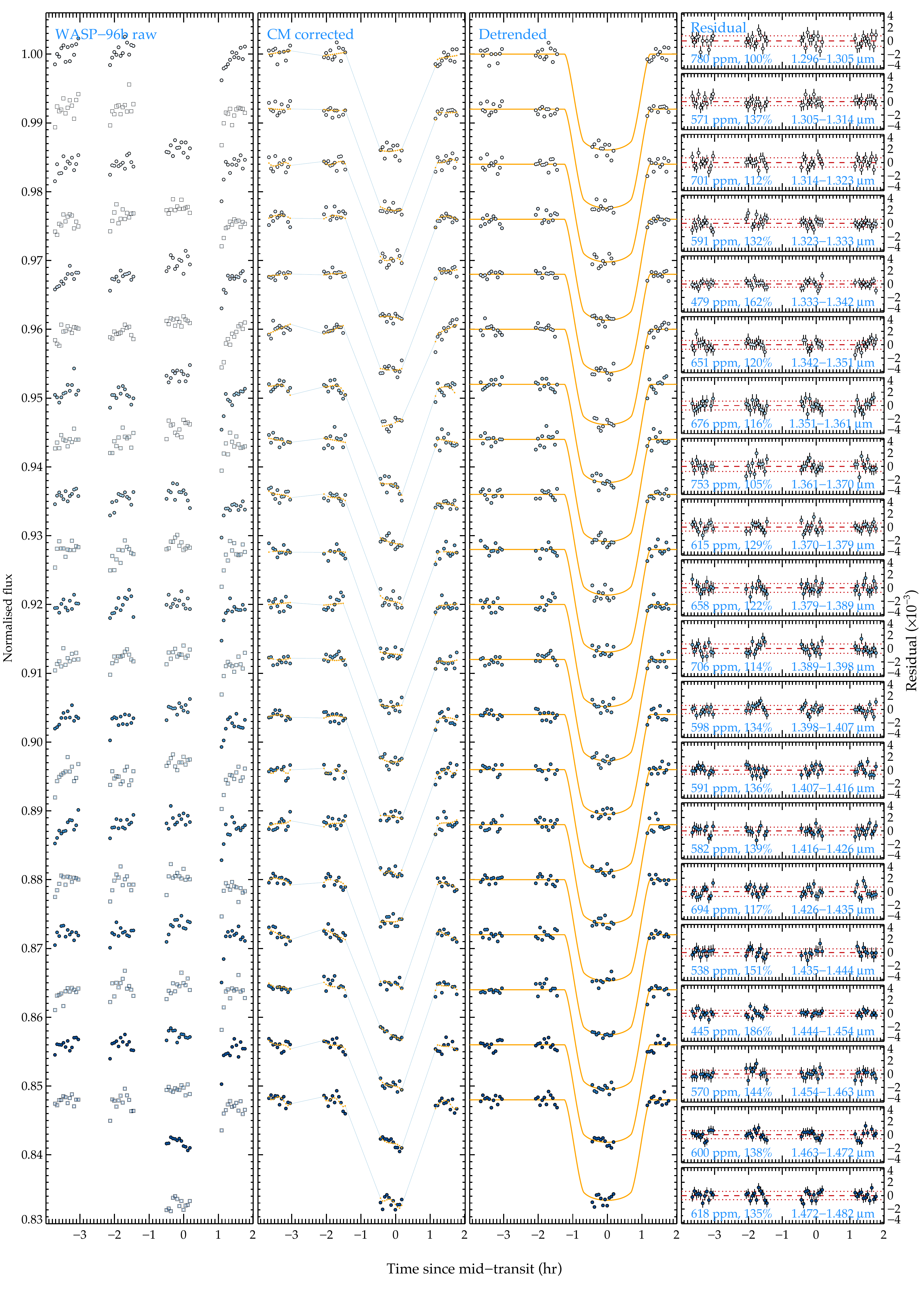}
    \caption{Same as Figure\ref{fig:figA4} but for grism G141. Shown is a subset of 20 light curves.}
    \label{fig:figA5}
\end{figure*}

\begin{figure*}
\centering
\includegraphics[trim={0.9cm 0.0cm 0.0cm 0.0cm}, width= 0.5\linewidth]{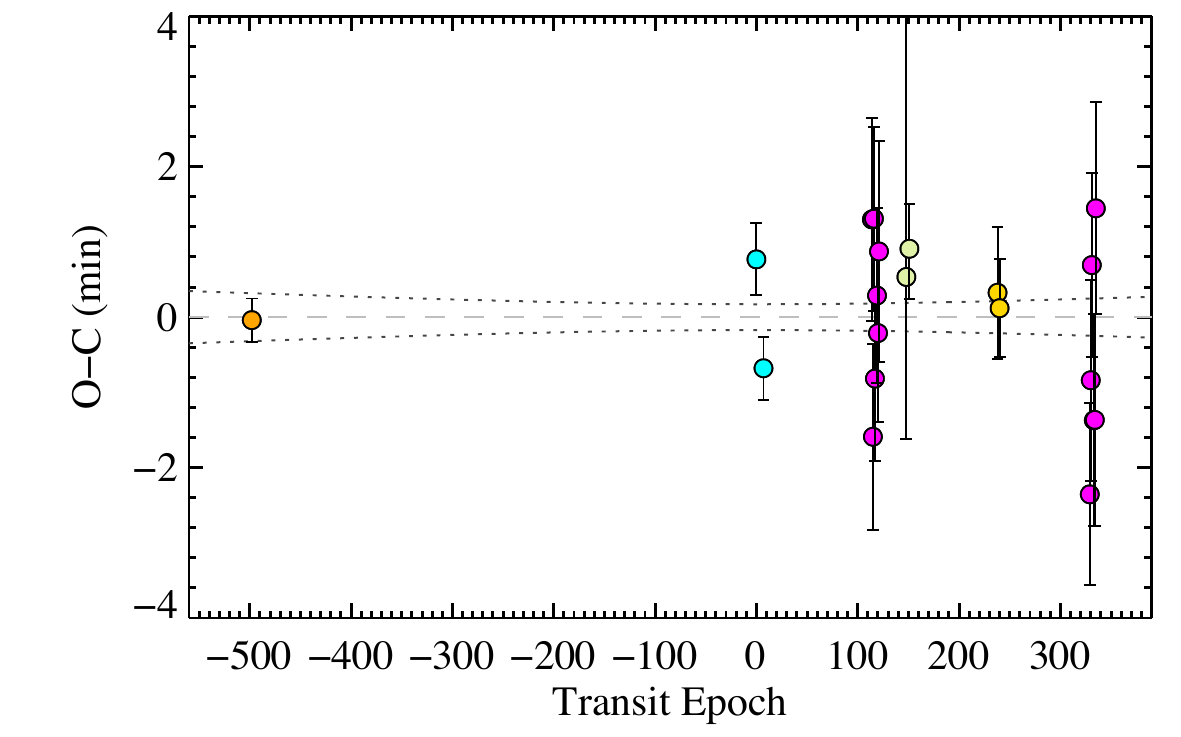}
\caption{Observed minus computed (O-C) transit times from the best-fitting orbital period and central transit times reported in the literature (orange: combined analysis of WASP, EulerCam and TRAPPIST light curves; cyan: VLT transits) and derived from our analysis of TESS (magenta), {\it{HST}} (green) and {\it{Spitzer}} (yellow) light curves. The error bars and dotted lines indicate the $1\sigma$ uncertainties of the transit times and derived ephemeris, respectively.}
\label{fig:figOC}
\end{figure*}

\section{Stellar activity and photometric variability}
This section contains figures from our analysis of the FEROS high-resolution spectroscopy and ASAS-SN light curve.

\begin{figure*}
\centering
\includegraphics[trim={0.9cm 0.0cm 0.0cm 0.0cm}, width= 0.5\linewidth]{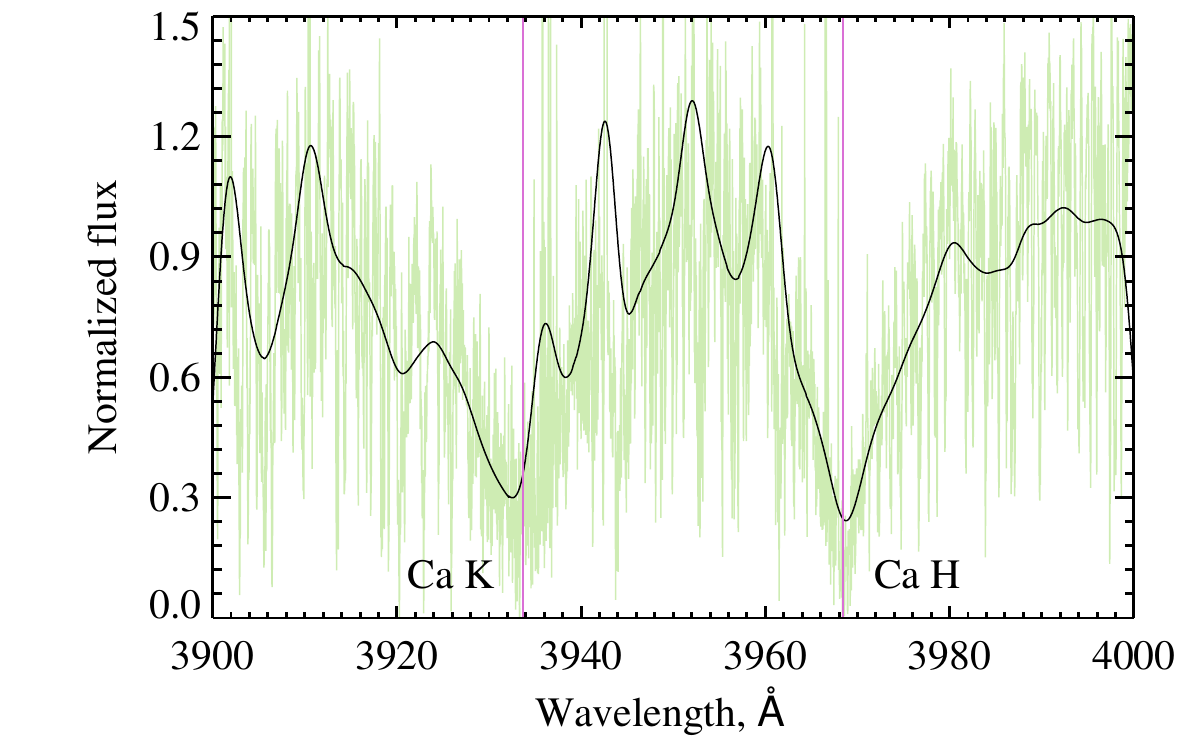}
\caption{FEROS spectroscopy around the Ca\,{\sc{II}}\,H\&K lines for WASP-96. The green line presents the original  data and the black line indicates a Gaussian smoothed version for clarity. The tabulated laboratory wavelengths of the  cores of the H$\&$K lines are plotted with the purple vertical lines.}
\label{fig:figHK}
\end{figure*}

\begin{figure*}
\centering
\includegraphics[trim={0.0cm 0.0cm 0.0cm 0.0cm}, width= 0.55\linewidth]{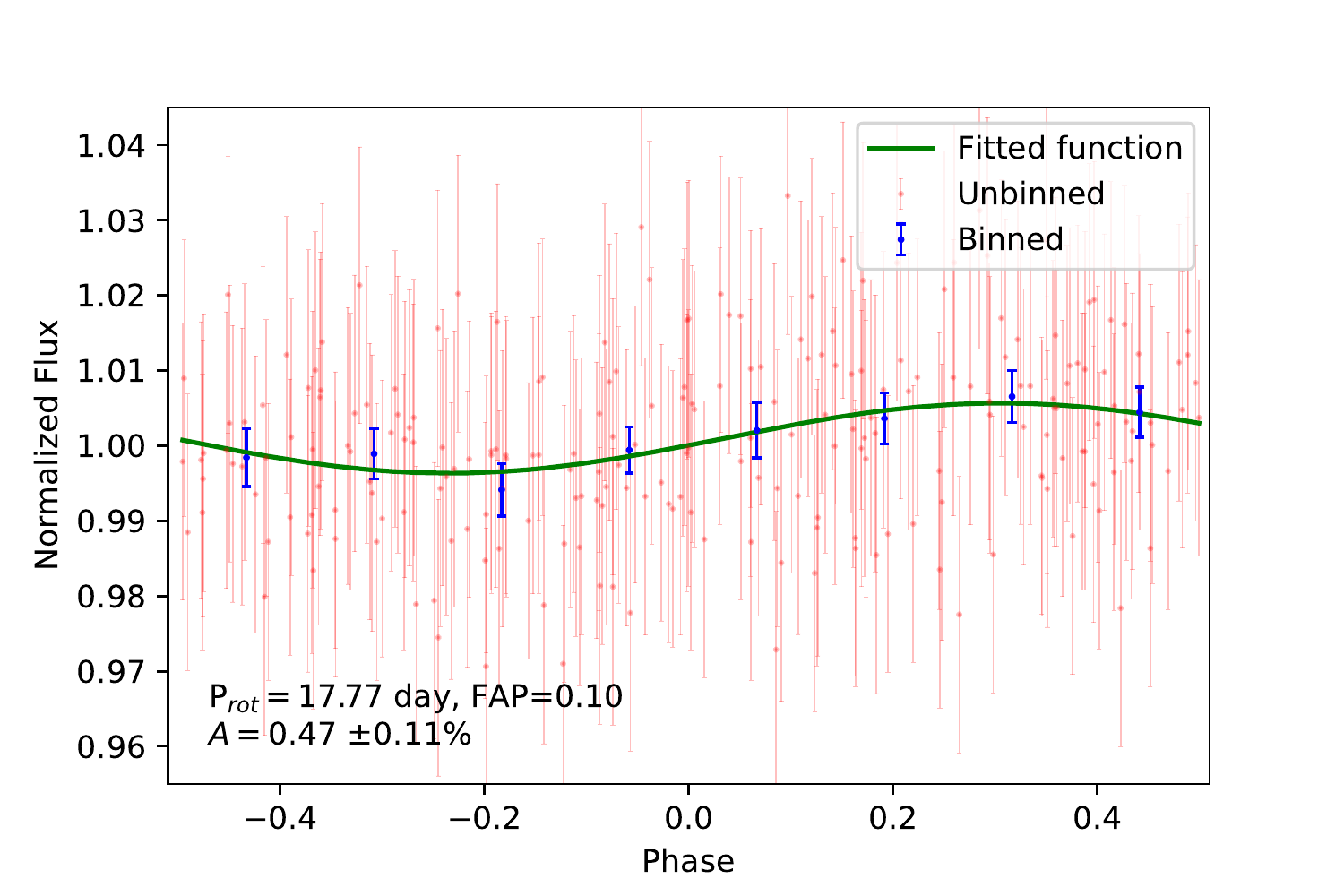}
\caption{Phase-folded photometry from the ASAS-SN observing campaigns with the period of highest significance, as determined in our periodogram analysis. Shown are the unbinned photometry (red), binned light curve (blue), and the best-fit sine function (green). The vertical bars indicate $1\sigma$ uncertainties.}
\label{fig:asassn_folded}
\end{figure*}

%\section{Stellar elemental abundances}
%This sections presents results about the elemental abundances of the planet host star measured form observations with the FEROS high-resolution spectrograph on the ESO/MPG 2.2m telescope.

%%%%%%%%%%%%%%%%%%%%%%%%%%%%%%%%%%%%%%%%%%%%%%%%%%

% Don't change these lines
\bsp	% typesetting comment
\label{lastpage}
\end{document}